\newif\ifdraftmode

\draftmodefalse

\ifdraftmode
\documentclass[journal, draftcls, onecolumn, 12pt]{IEEEtran}
\else
\documentclass[journal]{IEEEtran}
\fi

\usepackage{cite, xcolor}

\ifCLASSINFOpdf
  \usepackage[pdftex]{graphicx}
\else
\fi
\usepackage{amsmath, amsfonts, amssymb}
\usepackage{cases}
\usepackage{multirow}
\DeclareMathOperator*{\Tr}{Tr}

\DeclareMathOperator*{\diag}{diag}

\newtheorem{lemma}{Lemma}

\newtheorem{theorem}{Theorem}
\newtheorem{assumption}{Assumption}

\newtheorem{define}{Definition}

\usepackage{algorithmic}
\usepackage{algorithm}

\newcommand{\editrev}[1]{{\color{black}#1}}
\ifCLASSOPTIONcompsoc
\usepackage[caption=false,font=normalsize,labelfont=sf,textfont=sf]{subfig}
\else
\usepackage[caption=false,font=footnotesize]{subfig}
\fi
\usepackage{hyperref}
\begin{document}
%
\title{Uplink-Downlink Duality for Beamforming in Integrated Sensing and Communications}
%
%
%

\author{Kareem M. Attiah,~\IEEEmembership{Member,~IEEE,}
and~Wei~Yu,~\IEEEmembership{Fellow,~IEEE}
\thanks{Manuscript submitted to \emph{IEEE Journal on Selected Areas in
Information Theory} on September 16, 2025, and revised on November 25, 2025.
The authors are with the Electrical and Computer Engineering Department,
University of Toronto, Toronto, ON M5S3G4, Canada.  
Emails: kareemattiah@gmail.com, weiyu@ece.utoronto.ca.  The materials of this paper have been presented in part at the IEEE International Symposium on Information Theory (ISIT), Athens, Greece, July 2024~\cite{attiah2024uldl}. 
This work was supported by Natural Science and Engineering Research Council
(NSERC) via a Discovery Grant and via the Canada Research Chairs program.
}
}

\maketitle

\begin{abstract}

This paper considers the beamforming and power optimization problem for 
a class of integrated sensing and communications (ISAC) problems that utilize the communication signals simultaneously for sensing. We formulate the problem
of minimizing the Bayesian Cramér-Rao bound (BCRB) on the mean-squared error of 
estimating a vector of parameters, while satisfying downlink signal-to-interference-and-noise-ratio constraints for a set of communication
users at the same time. 
The proposed optimization framework comprises two key new ingredients. 
First, we show that the BCRB minimization problem corresponds to 
maximizing beamforming power along certain sensing directions of interest. 
Second, the classical uplink-downlink duality for multiple-input multiple-output communications can be extended to the ISAC setting, but unlike the classical communication problem, the dual uplink problem for ISAC 
may entail negative noise power and needs to include an extra condition on the
uplink beamformers. This new duality theory opens doors for efficient iterative 
algorithm for optimizing power and beamformers for ISAC. 


\end{abstract}
%

\IEEEpeerreviewmaketitle

\section{Introduction}

Traditional wireless communication systems focus on the task of transmitting 
information.  Although these traditional systems typically also perform
sensing-like tasks (e.g., channel estimation), the primary purpose of these
sensing operations is to enable or to enhance communications. In recent
years, there has been a growing interest in designing wireless systems that
are capable of delivering sensing functionalities separately from and in
addition to communications services \cite{liutcom202, liujsac2022}.
Such systems are expected to play a key role in emerging applications such as
autonomous driving, 
smart cities, and industrial automation.

Moreover, there is also an emerging trend to integrate sensing operations into
the communication system design. As opposed to implementing separate radar-like waveforms to perform sensing tasks independently of the communication tasks,
there is now an increasing interest in exploring whether it is possible to use
the same waveform to perform both sensing and communication tasks
simultaneously. This brings operational benefits such as shared utilization of
the wireless spectrum and hardware resources, as well as lower cost. 

This paper focuses on the task of designing common waveforms for integrated
sensing and communications (ISAC). This is a more challenging task than 
designing waveforms for communications alone, because the same 
information-bearing signals must also perform sensing. 
In the context of multiple-input multiple-output (MIMO) transmit beamforming, the beamformers that need to guarantee certain quality-of-service metrics for communication users must now also be optimized for the relevant performance 
metric for sensing \cite{Liu2020joint, HuaOptimal2023, LiuFCRB2022}. 


Specifically, this paper considers the problem of designing an optimal set of downlink beamformers 
that can be used by the base station to deliver optimal
sensing performance for estimating a vector of unknown parameters,
while \emph{simultaneously} satisfying a set of signal-to-interference-and-noise-ratio (SINR) constraints for the communications users. 

To quantify the sensing performance, we 
utilize the Cramér-Rao bound (CRB) as a metric for sensing. 
The optimization of CRB is not straightforward due to the fact that the structure of
the CRB involves a matrix inverse (e.g., see \cite{LiuFCRB2022, xu2024, Zhuinformation2023} for the scalar case where the CRB is a fraction).
One of the main contributions of this paper is that, under certain conditions,
the minimization of CRB can be transformed into a sequence of subproblems, 
each involving maximizing power in certain sensing directions of interest. 
This essentially reduces the ISAC problem to downlink 
beamforming problems that resemble those treated in classical works 
in MIMO communications
\cite{rashidUL1998,
rashidDL1998, Schubertsolution2004, yutransmitter2007, WieselLinear2006,
bengtsson2018optimum}. These subproblems are indexed by a set of auxiliary
variables, which can be optimized in an outer loop.



Furthermore, this paper shows that an uplink-downlink duality relationship 
can be developed for each of these subproblems, so that the optimal downlink beamformers can be obtained by solving a dual uplink problem.  But unlike the duality relationship in the classical MIMO communication setting 
\cite{rashidUL1998, rashidDL1998, Schubertsolution2004, yutransmitter2007, 
WieselLinear2006, bengtsson2018optimum}, the dual uplink problem for ISAC may entail negative noise power and involve extra conditions 
on the uplink beamformers. This paper provides a theoretical development of 
this new duality relationship, which can open doors for efficient numerical
algorithms for power and beamforming optimization for ISAC systems.



\subsection{Related Works}

Much existing work on ISAC has been devoted to the problem of beamforming design for simultaneous communications and sensing. In the existing literature, the beamforming design can take on one of many possible formulations: i) optimizing the sensing performance subject to a set of constraints on the user SINRs \cite{Liu2020joint, HuaOptimal2023, LiuFCRB2022, LiuF2022conf, attiahactive2023, wen2023, Chen2022}; ii) optimizing the user rates (e.g., sum or min rate) subject to a constraint on the sensing performance metric \cite{Chen2021, Chen2022, He2023}; or iii) optimizing a weighted average of the communication and sensing metrics \cite{Liu2018MUMIMO}. The question of which formulation is more appropriate depends on the specific application. In addition, the above works also differ based on the nature of sensing operation (i.e., detection versus estimation) and the adopted performance metric. 
Common metrics include beampattern-based designs \cite{Liu2020joint, HuaOptimal2023, Chen2021}, classical CRB \cite{LiuFCRB2022, LiuF2022conf}, Bayesian CRB (BCRB) \cite{xu2024}, radar signal-to-noise-ratio \cite{attiahactive2023,He2023}, radar SINR \cite{wen2023}, and radar signal-to-clutter-and-noise-ratio \cite{Chen2022}, etc.

In terms of beamformer design for ISAC, the works \cite{Liu2020joint, HuaOptimal2023, LiuFCRB2022, Chen2021, xu2024, attiahactive2023, Chen2022} all adopt the semidefinite relaxation (SDR) method, which relaxes the beamformer design problem into optimization of transmit covariance matrices. 
In addition to the high computational burden, these works assume that the set of beamformers can be extended to include additional beamformers dedicated to sensing only (see \cite{Liu2020joint} for details). Thus, these works do not exclusively rely on the communications beamformers for sensing, in contrast to the setting of the current paper. 

In a separate line of work, \cite{wen2023, He2023, LiuF2022conf} develop low-complexity iterative algorithms based on successive convex approximation (SCA), which allows the beamforming problem to be solved as a sequence of second-order cone programs (SOCPs). While these algorithms tend to be more efficient than the SDR methods, it is unknown whether they guarantee global convergence. It is worthwhile to remark that there are also other suboptimal low-complexity algorithms developed for different contexts, e.g., robust beamforming for ISAC \cite{choi2024joint}, and multi-cell ISAC \cite{chen2024fast}.

The above works primarily design the beamformer based on average statistics for communications. A recent line of work~\cite{lu2023random, Luoptimal2024} has considered data-dependent beamforming in order to mitigate the effect of randomness in the transmitted signal on the sensing performance. Although~\cite{lu2023random, Luoptimal2024} demonstrate that such a strategy may improve the sensing, it entails a more complicated optimization; also, the performance improvement diminishes at large block lengths.

This paper advocates the use of uplink-downlink duality for beamformer design.
Uplink-downlink duality is a beamforming design concept rooted in classical MIMO communications. The main idea first appears in~\cite{rashidDL1998}, which relies on transforming the difficult downlink communication problem into an equivalent virtual uplink problem~\cite{rashidUL1998}. The uplink-downlink duality concept has been instrumental in the development of capacity results for the MIMO downlink channel \cite{Viswanathsum2003, Vishwanathduality2003, yuminimax2006}. 
Moreover, the work \cite{yutransmitter2007} shows that this duality concept is intimately related to the Lagrange duality theory in optimization, and it can be further extended to solving the downlink beamforming problem with per-antenna power constraints. Since then, the duality principle has been adopted to tackle the downlink beamforming design in many contexts, including multi-cell coordinated beamforming \cite{Dahrouj2010,yang2008}, cooperative cellular systems \cite{ya-feng,Miretti2024}, and energy harvesting \cite{jie2014}.

\subsection{Main Contributions}
This paper develops an uplink-downlink duality relationship for solving an ISAC problem of minimizing CRB for sensing subject to SINR constraints for the communication users. The main contributions of this paper are as follows:
\begin{enumerate}

    \item 
	    To address the issue that the CRB has a complicated fractional structure, this paper introduces a set of auxiliary variables and shows that under certain conditions, the downlink ISAC problem can be transformed into a max-min formulation with a particular quadratic objective. Such a transformation not only yields an optimization problem in a considerably more tractable form, but also gives rise to a novel interpretation that optimizing the CRB is equivalent to the maximization of beamforming power along certain spatial directions.

    \item The classical duality relationship between the uplink and the downlink can be used to solve the downlink beamforming problem for minimizing the base station total transmit power subject to SINR constraints for the communication users. This paper shows that the ISAC beamforming problem can, in contrast, be regarded as a maximization of directional power, under a total power constraint and the SINR constraints. 
We develop a duality theory for this new setting, which allows the ISAC problem to be viewed as a virtual uplink communication problem with a modified noise covariance that encodes the sensing metric. 
Further, the uplink noise covariance is not necessarily positive semidefinite (PSD).
We show that duality continues to hold \emph{under an admissibility condition}. Moreover, the dual uplink problem entails an additional \emph{M-matrix} constraint on the beamformers. This new duality theory opens doors for efficient numerical algorithms for beamformer design in ISAC.

\end{enumerate}

\subsection{Paper Organization and Notation}

The rest of the paper is organized as follows. Section~\ref{sec:sysmdl} discusses the ISAC system model and presents the problem formulation. In Section~\ref{sec:pw_max}, we transform the ISAC problem into an equivalent max-min problem and provide interpretations of the new formulation.  Section~\ref{sec:duality} presents the uplink-downlink duality theory for the ISAC system. Section~\ref{sec:numerical} presents numerical results. The paper concludes in Section~\ref{sec:conc}. 

This paper uses lowercase, lowercase boldface, and uppercase boldface letters to denote scalars, vectors, and matrices, respectively. We use $(\cdot)^\textsf{T}$, $(\cdot)^\textsf{H}$, $(\cdot)^{-1}$, and $(\cdot)^\dagger$ to represent the transpose, Hermitian, inverse, and pseudo-inverse of a matrix. We use $\Re\{\cdot\}$, $\Im\{\cdot\}$ to denote the real and imaginary parts of a complex number, $\overline{(\cdot)}$ as the complex conjugate, and $\imath$ as $\sqrt{-1}$.
For vectors, $\| \cdot\|$ denotes the Euclidean norm. The matrix operators $\Tr(\cdot)$ and $\rho_\text{max}(\cdot)$ denote the trace and spectral radius. Note that for a PSD matrix, the spectral radius is its maximum eigenvalue. 
We use $\mathbf{I}$ to denote an identity matrix of appropriate dimensions with its $\ell$-th column denoted by $\mathbf{e}_\ell$. We use $\mathbf{1}$ to denote the vector of all ones and $\mathbf{0}$ a matrix (or vector) of all zeros with appropriate dimension. We use $\diag(d_1, \ldots, d_K)$ 
to denote a diagonal matrix with diagonal elements $d_1, \ldots, d_K$. 
The space of $n \times m$ matrices with real (complex) entries is denoted by $\mathbb{R}^{n \times m}$ ($\mathbb{C}^{n \times m}$).  The set of vectors in $\mathbb{R}^K$ with nonnegative (positive) entries is denoted by $\mathbb{R}^K_+$ ($\mathbb{R}^K_{++}$). 
We use $\mathbb{E}[\cdot]$ to denote the expectation of a random variable.
Finally, $\mathcal{CN}(\boldsymbol{\mu}, \boldsymbol{\Sigma})$ represents the distribution of circularly symmetric complex Gaussian vectors with mean $\boldsymbol{\mu}$ and covariance $\boldsymbol{\Sigma}$.


\section{System Model}
\label{sec:sysmdl}
\begin{figure}
    \centering    \includegraphics[width=0.45\textwidth]{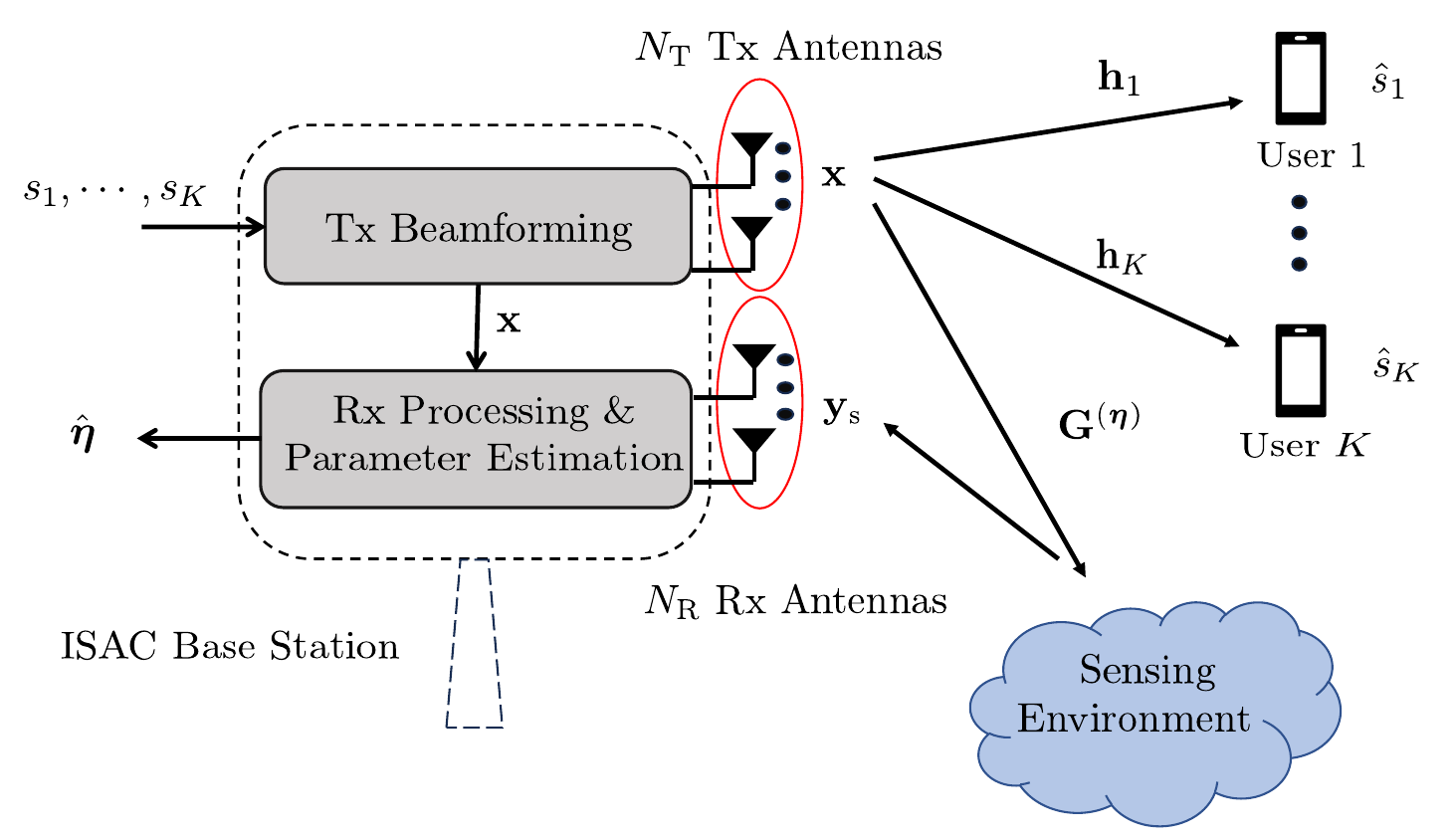}
	\caption{The ISAC system model where the BS serves $K$ communication users via linear beamforming while simultaneously aiming to learn some underlying parameter of the sensing environment using $N_\text{T}$ transmit (Tx) antennas and $N_\text{R}$ receive (Rx) antennas in full-duplex operation.}
     \label{fig:model}
\end{figure}

This paper considers a narrowband downlink multi-user ISAC system illustrated
in~\figurename~\ref{fig:model}, comprising a MIMO base station (BS) with $N_\text{T}$ transmit antennas and $N_\text{R}$ receive antennas. The BS provides two distinct functionalities: (i) downlink communications to $K$ single-antenna users, and (ii) monostatic sensing for estimating a vector of $L$ real-valued parameters denoted by $\boldsymbol{\eta} \in \mathbb{R}^{L}$. 

We study the ISAC operation over
a coherence interval spanning $T$ symbol periods in which both the sensing parameters and communication channels are assumed to be fixed.  In each symbol period, the BS transmits a narrowband waveform through its transmit array that conveys communication symbols to the remote users, while collecting the back-propagated echo signal through the receive array for sensing the parameters of interest. The communications and sensing take place at different time scales. The former happens during each symbol period, whereas the latter takes place over multiple symbols within the coherence interval. 

The parameters that can be estimated in this setting are essentially spatial parameters of the channel (e.g.,  angles of arrival in azimuth and elevation for multiple targets of interest) as opposed to temporal parameters (e.g., delay and Doppler)\footnote{The estimation of delay and Doppler requires either larger bandwidth (beyond the narrowband assumption) or longer observation windows (beyond the coherence interval), and is outside of the scope of the current investigation.}.



\subsection{Transmit Signal Model}

This paper adopts a signaling scheme in which the transmit vector across the antennas in the $t$-th symbol period, $\mathbf{x}[t] \in \mathbb{C}^{N_\text{T}}$, follows a linear beamforming model as the sum of $K$ beamformed communication symbols for the $K$ users: 
\begin{equation}
    \label{eq:BFMDL}
    \mathbf{x}[t] = \mathbf{V} \mathbf{s}[t] = \sum_{k = 1}^K \mathbf{v}_k s_k[t], \quad t = 1, \ldots, T,
\end{equation}
where $\mathbf{v}_k$ is the beamforming vector for the $k$-th user
and $\mathbf{s}[t] = \left[s_1[t], \ldots, s_K[t]\right]^\textsf{T} \in \mathbb{C}^{K}$ is a vector of communication symbols for the $K$ users modelled as a zero-mean i.i.d.\ Gaussian random vector with $\mathbb{E}[ \mathbf{s}[t] \mathbf{s}^\textsf{H}[t]] = \mathbf{I}$. 
We use $\mathbf{V} \triangleq \left[\mathbf{v}_1, \ldots, \mathbf{v}_K\right] \in \mathbb{C}^{N_\text{T} \times K}$ to denote the matrix of beamformers. 
A total power constraint $P$ is imposed on the transmit signal, i.e., we require $\Tr\left(\mathbf{V} \mathbf{V}^\mathsf{H}\right) \leq P$.

The beamforming model \eqref{eq:BFMDL} is exactly the same as that of
conventional communication systems in that the total number of beamformers is equal to the number of users. The key consideration here, however, is that the communication symbols are also used for sensing, i.e., these beamformers are designed to optimize both the sensing performance and the communication performance.

The above beamforming model can be contrasted with the so-called extended model considered in many existing ISAC studies, e.g.,~\cite{Liu2020joint,HuaOptimal2023, LiuFCRB2022, Zhuinformation2023}, where the $K$ beamformers are augmented with \emph{additional} beamformers for sensing, i.e.,
\begin{equation}
    \label{eq:BFMDL_extended}
    \mathbf{\tilde x}[t] = \mathbf{V} \mathbf{s}[t] + \mathbf{V}_\text{s} \mathbf{s}_\text{s}[t].
\end{equation}
where $\mathbf{V}_\text{s} \in \mathbb{C}^{N_\text{T} \times N_\text{T}}$ is a matrix of additional beamformers (referred to as the sensing beamformers) and $\mathbf{s}_\text{s}[t] \in \mathbb{C}^{N_\text{T}}$ is typically a known pseudo-random sequence with $\mathbb{E}[ \mathbf{s}_\text{s}[t] \mathbf{s}^\textsf{H}_\text{s}[t]] = \mathbf{I}$. This latter model can potentially enhance the sensing performance as compared to the model in (\ref{eq:BFMDL}), since it offers additional degrees of freedom, although it is also more complex to implement. 

However, there are many scenarios for which the model in \eqref{eq:BFMDL}
already performs as well as the extended model. For these systems, the model
\eqref{eq:BFMDL} is much more desirable than the extended model, since the
model \eqref{eq:BFMDL} requires no modifications to the existing communication
transmission scheme.  This paper restricts attention to these scenarios
and studies beamformer design strategies for the class of ISAC systems where 
the model \eqref{eq:BFMDL} already provides the optimal sensing performance. 
We will define this class of ISAC systems more precisely and give sufficient
conditions for these systems in a subsequent section, after first describing
the respective channel models and performance metrics for sensing and
communications.

\subsection{Sensing Model and Performance Metric}

The sensing model adopted in this paper assumes that the BS can perform self-interference mitigation~\cite{Sabhinband2014}, i.e., the leakage from the transmit array to the received antennas can be cancelled. Consequently,  the baseband signal signal $\mathbf{y}_\text{s}[t] \in \mathbb{C}^{N_\text{R}}$ received at the BS for the $t$-th symbol \editrev{(after removing the effect of channel delay)} can be written as 
\begin{equation}
    \label{eq:radarmdl}
    \mathbf{y}_\text{s}[t] = 
    \mathbf{G}^{\left(\boldsymbol\eta \right)} 
    \mathbf{x}[t] + \mathbf{n}_\text{s}[t], \quad t = 1, \ldots, T
\end{equation}
where $\mathbf{G}^{\left(\boldsymbol\eta \right)}  \in \mathbb{C}^{N_\text{R} \times N_\text{T}}$ is the sensing channel matrix that models the ``round-trip-return'' channel between the BS transmit and receive antenna arrays, and $\mathbf{n}_\text{s}[t] \sim \mathcal{CN}(0, \sigma_\text{s}^2 \mathbf{I})$ is the additive noise at the receiver. 

Observe that the sensing channel is parameterized by the parameters of interest $\boldsymbol{\eta}$. Throughout this paper, we do not specify the exact parametric dependency between $\mathbf{G}^{\left(\boldsymbol\eta \right)}$ and $\boldsymbol{\eta}$. We only make the assumptions that
\begin{itemize}
\item $\mathbf{G}^{\left(\boldsymbol\eta \right)}$ is a deterministic function of $\boldsymbol{\eta}$, and 
\item the functional form of $\mathbf{G}^{\left(\boldsymbol\eta \right)}$ is known.
\end{itemize}
These are reasonable assumptions as the parametric form $\mathbf{G}^{\left(\boldsymbol\eta \right)}$ can typically be obtained through appropriate channel modelling. 

The goal for sensing is to estimate $\boldsymbol{\eta}$. 
The above model implicitly assumes that it is of interest to
estimate all the parameters of the channel $\mathbf{G}^{\left(\boldsymbol\eta \right)}$. 
The scenario in which the sensing objective is to estimate only some part of
$\boldsymbol{\eta}$ can be handled by incorporating a weighting factor for each
component of $\boldsymbol{\eta}$ in the sensing objective, and then setting
the weights corresponding to the ``nuisance'' parameters of no interest to zero, 
as will be explained in more detail shortly. In this way, the above formulation can 
encompass a wide variety of sensing problems, including, e.g., estimating 
angles-of-arrival of multiple targets or target position in the near-field.

This work adopts a Bayesian estimation framework by assuming the availability
of a prior distribution $f(\boldsymbol{\eta})$ and uses the BCRB to quantify
the sensing performance for estimating $\boldsymbol{\eta}$. 
The prior distribution $f(\boldsymbol{\eta})$ can be obtained, for example, 
from historical measurements in previous coherence intervals.

Given the input to the channel $\mathbf{X} \triangleq \left[\mathbf{x}[1], \ldots, \mathbf{x}[T]\right]$ and the observed output $\mathbf{Y}_\text{s} \triangleq [\mathbf{y}_\text{s}[1], \ldots, \mathbf{y}_\text{s}[T]]$ within the coherence interval,
let  $\hat{\boldsymbol{\eta}} \triangleq \hat{\boldsymbol{\eta}}\left(\mathbf{X}, \mathbf{Y}_\text{s}\right)$ denote any well-behaved
estimator of $\boldsymbol{\eta}$ \cite{van2004detection}.
The BCRB states that 
\begin{equation} \label{eq:crb}
	\mathbb{E}_{\boldsymbol \eta} \left[ \mathbb{E}_{\mathbf{X}, \mathbf{Y}_\text{s}} \left[ \left( \boldsymbol{\eta} - \hat{\boldsymbol{\eta}} \right) \left( \boldsymbol{\eta} - \hat{\boldsymbol{\eta}}\right)^\textsf{T} \big{|} \boldsymbol{\eta} \right]  \right] \succcurlyeq \mathbf{J}^{-1}_\mathbf{V}, 
\end{equation}
where $\succcurlyeq$ denotes inequality with respect to the PSD cone, and the inner expectation\footnote{Note that we take the average over $\mathbf{X}$ since it is a random communication signal carrying information intended for the communication users.} is for a fixed $\boldsymbol{\eta}$ and is taken jointly over all possible realizations of $\mathbf{X}$, $\mathbf{Y}_\text{s}$, and the outer expectation is over the prior $f(\boldsymbol{\eta})$. 

The PSD matrix $\mathbf{J}_\mathbf{V}$ is the $L \times L$ Bayesian Fisher information matrix (BFIM), whose $(i, j)$-th element is \cite{van2004detection}
    \begin{align}    \left[\mathbf{J}_\mathbf{V} \right]_{ij} &\triangleq -\mathbb{E}_{\mathbf{X}, \mathbf{Y}_\text{s},\boldsymbol{\eta}}  \left[ \frac{\partial^2 \log{ f\left(\mathbf{X}, \mathbf{Y}_\text{s}, \boldsymbol{\eta} \right) } }{\partial \eta_i \partial \eta_j} \right]. 
    \end{align}
The subscript $\mathbf{V}$ here highlights the fact that the BFIM depends on the input 
distribution only through the beamforming matrix $\mathbf{V}$.

It can be shown \cite{attiah2025howmany} that for the considered model~\eqref{eq:BFMDL}-\eqref{eq:radarmdl} with the assumption that $\mathbf{G}^{(\boldsymbol{\eta})}$ is a deterministic function of $\boldsymbol{\eta}$, the BFIM can be expressed as the sum of two components
    \begin{equation} \label{eq:FIM_elements}
    \mathbf{J}_\mathbf{V} =   \mathbf{C} + \mathbf{T}_\mathbf{V},
    \end{equation}
where $\mathbf{C} \in \mathbb{R}^{L \times L}$ is a PSD matrix that depends only on the prior distribution, and $\mathbf{T}_\mathbf{V} \in \mathbb{R}^{L \times L}$ is a PSD matrix that captures the dependency on the beamforming matrix $\mathbf{V}$.  

Let $\dot{\mathbf{G}}_i^{(\boldsymbol{\eta})} \triangleq \tfrac{\partial \mathbf{G}^{\left(\boldsymbol\eta \right)} }{\partial \eta_i}$. The $(i, j)$-th elements of these two matrices are given by
\begin{equation}~\label{eq:c_elements}
	[\mathbf{C}]_{ij} = - \mathbb{E}_{\boldsymbol{\eta}} \left[\frac{\partial^2 \log f(\boldsymbol{\eta}) }{\partial \eta_i \partial \eta_j}\right],
\end{equation}
and
\begin{equation}~\label{eq:d_elements}
	[\mathbf{T}_\mathbf{V}]_{ij}  =  \mathbb{E}_{\boldsymbol{\eta}} \left[\frac{T}{\sigma_\text{s}^2}  \Tr\left( \ddot{\mathbf{G}}_{ij}^{(\boldsymbol{\eta})}  \mathbf{V} \mathbf{V}^\textsf{H} \right)\right],
\end{equation}
where $\ddot{\mathbf G}_{ij}^{(\boldsymbol{\eta})} \triangleq {(\dot{\mathbf{G}}^{(\boldsymbol{\eta})}_i)}^\textsf{H} \dot{\mathbf{G}}^{(\boldsymbol{\eta})}_j + {(\dot{\mathbf{G}}^{(\boldsymbol{\eta})}_j)}^\textsf{H} \dot{\mathbf{G}}^{(\boldsymbol{\eta})}_i$. 

Asymptotically, either as $\sigma_\text{s}^2 \rightarrow 0$ or as $T \rightarrow \infty$,
the BCRB provides a tight bound on the mean-squared 
error (MSE) matrix associated with the optimal (i.e., the minimum mean-squared error (MMSE)) estimator. 

Given the BFIM, one can define different possible scalar functions for measuring the sensing performance, e.g., the trace-inverse of BFIM, or the logarithm of the determinant of BFIM. In this work, we consider a weighted-sum criterion
\begin{equation} \label{eq:sensing_metric}
    \mu(\mathbf{V}) \triangleq \Tr \left(\mathbf{W} \mathbf{J}^{-1}_ \mathbf{V}\right)
\end{equation}
where $\mathbf{W} = \diag \left(w_1, \ldots, w_L\right)$ is a matrix whose diagonal elements indicate the relative importance for estimating the different parameters in $\boldsymbol{\eta}$.
For the case where only a subset of the parameters in $\boldsymbol{\eta}$ are of interest, 
the weights corresponding to the nuisance parameters can be set to be zero.

We remark that the BCRB considered in this work coincides with the notion in standard textbook in estimation (e.g., \cite{van2004detection}) that treats $(\mathbf{X}, \mathbf{Y}_\text{s})$ as random observations revealed to the estimator. In the literature, tighter bounds on the MSE are also known, e.g., the Weiss-Weinstein bound~\cite{weiss1985} and the Ziv-Zakai bound~\cite{ziv1969}. Furthermore, even within the CRB family, the so-called Miller-Chang bound~\cite{miller1978} is reported to be tighter in some scenarios. 
Instead of setting the weights corresponding to the nuisance parameters to be zero, the Miller-Chang technique \cite{miller1978} takes statistical average over the nuisance parameters in the computation of CRB.
However, these potentially tighter bounds are also more difficult to compute and to optimize in most cases. 
In this paper, we restrict attention to the BCRB as given by the BFIM in~\eqref{eq:FIM_elements}-\eqref{eq:d_elements}.

The main advantage of using the BCRB as the sensing metric is that the BFIM has a closed-form that is relatively simple to compute as compared to the exact MSE calculation. Furthermore, it abstracts the receiver design since the bound holds for 
all practical estimators. Further, we note that unlike the classical CRB used in previous works~\cite{LiuFCRB2022, LiuF2022conf}, the BCRB does not depend on the actual value of the unknown parameter $\boldsymbol{\eta}$ and only the prior $f(\boldsymbol{\eta})$.

\editrev{We remark that the preceding formulation can be applied to the setting of active sensing \cite{foad_active}, where sensing occurs adaptively across multiple stages of the coherence interval, by updating the prior from one stage to the next as the BS learns more about the parameter \cite{attiahactive2023}. This can improve the sensing performance, but at the cost of increased computational complexity of optimizing the sensing strategy in each stage.}

\subsection{Communication Model and Performance Metric}
From the prespective of communications, upon transmitting $\mathbf{x}[t]$, each remote user 
receives a scalar baseband signal, denoted by $y_k[t]$, as given by
\begin{equation}
        \label{eq:commmdl}
    y_k[t] =
    \mathbf{h}_k^\textsf{H}
    \mathbf{x}[t] + n_k[t], \quad \forall k = 1, \ldots, K,
\end{equation}
where $\mathbf{h}_k \in \mathbb{C}^{N_\text{T}}$ is the communication channel of the $k$-th user, and $n_k[t] \sim \mathcal{CN}\left(0, \sigma_\text{c}^2\right)$ is the additive white Gaussian noise. 
In this paper, we make the assumption that the communication channel $\mathbf{H} \triangleq \left[\mathbf{h}_1, \ldots, \mathbf{h}_K\right] \in \mathbb{C}^{N_\text{T} \times K}$ can be perfectly estimated within each coherence interval (e.g., using uplink pilots and assuming channel reciprocity), and is known perfectly at the BS. 

The goal of communications is to satisfy certain achievable rate requirements for the users, which translate to constraints on the SINR at the communication receivers. Specifically, for each of the users $k= 1, \ldots, K$, we impose the following SINR 
constraint 
\begin{equation}
    \label{eq:SINR}
    \text{SINR}^{\text{DL}}_k(\mathbf{V}) \triangleq 
\frac{\left|\mathbf{h}_k^\textsf{H} \mathbf{v}_{k} \right|^2}{\sum_{i \neq k} \left| \mathbf{h}_k^\textsf{H} \mathbf{v}_{i}\right|^2 + \sigma_\text{c}^2} \geq \gamma_k,
\end{equation}
where $\gamma_k$ denote the SINR threshold for the $k$-th user.
Throughout this paper, we assume that $\gamma_k$ is positive for all $k$ (i.e., the users have positive rate constraints), and that the SINR constraints are strictly feasible.

\subsection{Problem Formulation}\label{sec:reformulation}

Based on the communications and sensing performance metrics discussed in the previous section, the beamforming design problem for joint sensing and communications can be formulated as that of optimizing the sensing performance subject to constraints on the communication SINRs within each coherence interval, i.e.,
\begin{subequations}\label{prob:generalCase}
    \begin{align}
    \underset{\mathbf{V}}{\mathrm{minimize}} ~~~~ & \Tr \left(\mathbf{W} \mathbf{J}^{-1}_\mathbf{V} \right) \label{eq:general_obj} \\
            \mathrm{subject \ to}  ~~~ & \frac{\left|\mathbf{h}_k^\textsf{H} \mathbf{v}_k \right|^2}{\sum_{i \neq k} \left|\mathbf{h}_k^\textsf{H} \mathbf{v}_i \right|^2 + \sigma_\text{c}^2} \geq \gamma_k, \quad \forall k, \label{eq:SINR_const} \\
            & \Tr\left( \mathbf{V} \mathbf{V}^\mathsf{H} \right) \leq P. \label{eq:pw_const}
    \end{align}
\end{subequations}
Problem~\eqref{prob:generalCase} is challenging to solve because
the objective has a complicated fractional form, which is difficult to optimize as a function of the beamforming matrix $\mathbf V$. Furthermore, neither the objective nor the feasible set is convex in $\mathbf{V}$. 

The main goal of this paper is to show that under the following key assumption, the problem~\eqref{prob:generalCase} can be solved efficiently using an uplink-downlink duality technique.

\begin{assumption}\label{main_assume}
Throughout this paper, we assume that the ISAC problem with 
the communication channel $\mathbf{H}$, the sensing channel $\mathbf{G}^{(\boldsymbol{\eta})}$, the prior $f(\boldsymbol{\eta})$, and power constraint $P$, has the following property. 
Let $\mu \in \mathbb{R}_+$ denote the trace of weighted BCRB, and let $\boldsymbol{\gamma} \in \mathbb{R}^K_{++}$ denote a positive vector of downlink communication SINR thresholds. 
We assume that whenever $(\mu, \boldsymbol{\gamma})$ is achievable using the extended model~\eqref{eq:BFMDL_extended}, i.e., there exists an extended matrix $\mathbf{V}_\text{ex} = \left[\begin{matrix} \mathbf{V} & \mathbf{V}_\text{s}  \end{matrix} \right]$ with a communication beamforming matrix  $\mathbf{V} = \left[ \mathbf{v}_1, \ldots, \mathbf{v}_K \right] \in \mathbb{C}^{N_\text{T} \times K}$ and a sensing beamformer matrix $\mathbf{V}_\text{s} \in \mathbf{C}^{N_\text{T} \times N_\text{T}}$ achieving 
\begin{equation}
    \mu = \Tr(\mathbf{W} \mathbf{J}^{-1}_{\mathbf{V}_\text{ex}}), 
\end{equation}
and
\begin{equation} 
\frac{\left|\mathbf{h}_k^\textsf{H} \mathbf{v}_{k} \right|^2}{\sum_{i \neq k} \left| \mathbf{h}_k^\textsf{H} \mathbf{v}_{i}\right|^2 + \mathbf{h}_k^\mathsf{H} \mathbf{V}_\text{s} \mathbf{V}_\text{s}^\mathsf{H} \mathbf{h}_k + \sigma_\text{c}^2} \geq \gamma_k, \quad \forall k,
\end{equation}
the same pair $(\mu, \boldsymbol{\gamma})$ is also achievable using $K$ beamformers alone under the same power constraint $P$, i.e., using model~\eqref{eq:BFMDL} with some $\mathbf{V} = \left[ \mathbf{v}_1, \ldots, \mathbf{v}_K \right] \in \mathbb{C}^{N_\text{T} \times K}$. 
\end{assumption}

In the extended model, the extra sensing beams $\mathbf{V}_\text{s}$ may help improve the sensing performance, but they also consume additional power and cause extra interference to the communication users. 
Assumption~\ref{main_assume} states that $\mathbf{V}_\text{s} = \mathbf{0}$
already achieves optimal performance, in which case 
the model reduces to \eqref{eq:BFMDL}.



The preceding assumption may seem restrictive at first glance, but recent studies have established several sufficient conditions under which Assumption~\ref{main_assume} holds. For instance, as demonstrated in \cite{attiah2025howmany}, when the number of users $K$ is large as compared to the number of parameters to be estimated $L$, more precisely, when  
\begin{equation}\label{eq:sufficient_condition}
    K \geq \frac{L (L + 1)}{4},
\end{equation}
using $K$ beamformers is sufficient for achieving the performance of the extended beamforming model \emph{regardless} of system parameters such as $N_\text{T}$, $N_\text{R}$, and channel realization. Similar conditions have also been established in several 
other studies~\cite{mateen2023, salman, yao2025optimal} for some special cases. 

It is important to note that most of these conditions are only sufficient 
conditions. For instance, even when $K < \tfrac{L (L + 1)}{4}$, 
Assumption~\ref{main_assume} may still hold, depending on the number of transmit 
and receive antennas, and the particular realizations of the communication channels,
the sensing channel, the prior, and the total power constraint. 


\subsection{Example} 

Consider an
example of estimating the angle-of-departure (AoD) and the angle-of-arrival
(AoA) for a target located in the far field of a full-duplex MIMO transceiver
equipped with $N_\text{T} = N_\text{R}$ antennas in a monostatic setting. 
The sensing channel is assumed to have a single line-of-sight (LoS) path so the
AoA is the same as the AoD, and the channel can be modelled as 
\begin{equation}
\mathbf{G}^{(\boldsymbol{\eta})}_{\text{AoA}} = \alpha \mathbf{A}(\theta),
\end{equation}
where $\alpha \in \mathbb{C}$ is the path loss coefficient and $\theta \in \mathbb{R}$
is the AoA/AoD, and $\mathbf{A}(\cdot) \in \mathbb{C}^{N_\text{T} \times N_\text{R}}$ is the combined array response given by
\begin{equation}
	\mathbf{A}(\theta) = \mathbf{a}_\text{R}(\theta) \mathbf{a}_\text{T}^\textsf{H}(\theta).
\end{equation}
For a uniform linear array of $N_\text{T}$ transmit antennas, we have
\begin{equation*}
\mathbf{a}_\text{T}(\theta) = \frac{1}{\sqrt{N_\text{T}}}  \left[e^{\imath \pi 0 \sin(\theta)}, \ldots, e^{\imath \pi \left(N_\text{T} - 1\right) \sin(\theta)} \right]^\textsf{T}. 
\end{equation*}
The receive array response $\mathbf{a}_\text{R}(\theta)$ is defined similarly. 

The set of unknown parameters $\boldsymbol{\eta} = [\Re\{\alpha\}, \Im\{\alpha\}, \theta]^\textsf{T}$ describes $\mathbf{G}^{(\boldsymbol{\eta})}$ in a deterministic fashion. 
The goal here is to estimate $\theta$; the path loss $\alpha$ is the nuisance parameter. 

Now, assuming a Gaussian prior for $\alpha$ and $\theta$ with variances $\tfrac{\sigma_\alpha^2}{2T} \sigma_\text{s}^2$ and $\tfrac{\sigma_\theta^2}{2T} \sigma_\text{s}^2$, respectively, and arbitrary means, the BFIM in~\eqref{eq:FIM_elements} can be shown to be
\begin{equation}\label{eq:BFIM_AoA}
\mathbf{J}_\mathbf{V} = \frac{2T}{\sigma_\text{s}^2} \left[ \begin{matrix} 
        h_\alpha(\mathbf{V}) +  \frac{1}{\sigma_\alpha^2}        & 0             & \Re\{h_{\theta\alpha}(\mathbf{V})\} \\
        0           & h_{\alpha}(\mathbf{V})  +  \frac{1}{\sigma_\alpha^2}         & \Im\{h_{\theta\alpha}(\mathbf{V})\} \\
        \Re\{h_{\theta\alpha}(\mathbf{V})\}  & \Im\{h_{\theta\alpha}(\mathbf{V})\}    & h_\theta(\mathbf{V}) +  \frac{1}{\sigma_\theta^2}
    \end{matrix} \right]
\end{equation}
where 
\begin{eqnarray}
h_\alpha(\mathbf{V}) & \triangleq & \Tr\left(\mathbb{E}_\theta[ \mathbf{A}^\textsf{H}(\theta) \mathbf{A}(\theta) ] \mathbf{V} \mathbf{V}^\textsf{H} \right), \\ 
h_{\theta \alpha}(\mathbf{V}) & \triangleq & \Tr\left(\mathbb{E}_{\alpha, \theta}[\alpha \mathbf{A}^\textsf{H} (\theta)\dot{\mathbf{A}}(\theta)] \mathbf{V} \mathbf{V}^\textsf{H} \right), \\ 
h_\theta(\mathbf{V}) & \triangleq & \Tr\left(\mathbb{E}_{\alpha, \theta}[|\alpha|^2 \dot{\mathbf{A}^\textsf{H}}(\theta) \dot{\mathbf{A}}(\theta)] \mathbf{V} \mathbf{V}^\textsf{H} \right),
\end{eqnarray} 
and $\dot{\mathbf{A}}(\theta) \triangleq \tfrac{\partial \mathbf{A}(\theta)}{\partial \theta}$. 
Since the goal is to estimate $\theta$ (but not $\alpha$), we set the weights to be $\mathbf{W} = \diag(0, 0, 1)$, so that $\Tr\left(\mathbf{W} \mathbf{J}_\mathbf{V}^{-1}\right) = \left[\mathbf{J}_\mathbf{V}^{-1}\right]_{3, 3}$. By utilizing the Schur complement relation 
\begin{equation}
\left[\mathbf{J}^{-1}\right]_{3, 3} = (s - \mathbf{r}^\mathsf{H} \mathbf{S}^{-1} \mathbf{r} )^{-1}, 
\end{equation}
for a $3 \times 3$ matrix $\mathbf{J} = \left[ \begin{matrix}
    \mathbf{S} & \mathbf{r} \\ \mathbf{r}^\mathsf{H} & s
\end{matrix} \right]$ with $\mathbf{S} \in \mathbb{R}^{2 \times 2}, \mathbf{r} \in \mathbb{R}^{2 \times 1}$ and $s \in \mathbb{R}$, 
it can be shown \cite{bekkerTarget2006} that problem~\eqref{prob:generalCase} reduces to 
\begin{align}\label{prob:simpleCase}
\underset{\mathbf{V} \in \mathcal{V}}{\mathrm{maximize}} ~~& h_\theta(\mathbf{V}) - \frac{\left|h_{\theta\alpha}(\mathbf{V}) \right|^2}{h_\alpha(\mathbf{V}) + \tfrac{1}{\sigma_\alpha^2}},
\end{align}
where $\mathcal{V}$ denotes the communication constraints~\eqref{eq:SINR_const}-\eqref{eq:pw_const}. 

As this example shows, the BCRB minimization problem is a complicated 
nonconvex optimization problem with a fractional 
structure. This paper aims to show that under Assumption~\ref{main_assume},
this problem can be solved efficiently. We now discuss the conditions
under which Assumption~\ref{main_assume} holds.

For the above example, since $L=3$, the sufficient condition~\eqref{eq:sufficient_condition} for Assumption~\ref{main_assume} to hold is $K \geq 3$. But in
fact, an improved version of this condition can be derived based on 
a result in \cite{attiah2025howmany}, which states $K  \geq \frac{d}{2}$ is already sufficient,
where $d$ is the number of distinct quadratic terms in the sensing objective. 
In problem~\eqref{prob:simpleCase}, we have $d = 4$, since $h_\theta(\cdot)$ and $h_\alpha(\cdot)$ both contain one quadratic term each, and $h_{\theta\alpha}(\cdot)$ contains two, namely 
\begin{align*}
    \Re\{h_{\theta \alpha}(\mathbf{V})\}  &= \Tr\left( \mathbb{E}_{\alpha, \theta}[\alpha \mathbf{A}^\textsf{H}(\theta)\dot{\mathbf{A}}(\theta) + \overline{\alpha} \dot{\mathbf{A}}^\mathsf{H}(\theta) \mathbf{A}(\theta) ]  \mathbf{V} \mathbf{V}^\textsf{H} \right) \\
    \Im\{h_{\theta \alpha}(\mathbf{V})\}  &= \Tr\left( \imath \mathbb{E}_{\alpha, \theta}[\alpha \mathbf{A}^\textsf{H}(\theta) \dot{\mathbf{A}}(\theta) - \overline{\alpha} \dot{\mathbf{A}}^\mathsf{H}(\theta) \mathbf{A}(\theta) ]  \mathbf{V} \mathbf{V}^\textsf{H} \right)
\end{align*} 
Note that quadratic terms must be of the form $\Tr\left(\mathbf{Q}
\mathbf{V}\mathbf{V}^\mathsf{H}\right)$ where $\mathbf{Q}$ is a Hermitian matrix. 
Thus, the real and imaginary parts of $h_{\theta\alpha}(\mathbf{V})$ need to 
be counted separately; see~\cite{attiah2025howmany} for details. With $d = 4$,
a sufficient condition for Assumption~\ref{main_assume} to hold for this example
of AoA estimation is simply
\begin{equation}
K \ge 2.  \label{eq:s_condition_v2}
\end{equation}
Thus, for an ISAC system serving at least two communication users, while estimating an
AoA at the same time, Assumption~\ref{main_assume} holds, i.e., using only 
communication beams can already achieve the performance of 
the extended model with extra sensing beams. 

We remark that even when $K = 1$, Assumption~\ref{main_assume} may still hold depending on the system parameters (e.g.,  $N_\text{T}$, $N_\text{R}$, channel realization, and the prior, etc.)
For instance, if $\alpha$ and $\theta$ are independent with $\mathbb{E}\left[ \alpha \right] = 0$, the problem of estimating $\theta$ in~\eqref{prob:simpleCase} reduces to
\begin{align}
\underset{\mathbf{V} \in \mathcal{V}}{\mathrm{maximize}} ~~& h_\theta(\mathbf{V}) 
\end{align}
which has a single quadratic term in the objective, as opposed to four for the general case. Using $d = 1$ in the condition in \cite{attiah2025howmany} yields $K \geq \frac{d}{2} = \tfrac{1}{2}$. This means that Assumption~\ref{main_assume} holds for this special case when $K = 1$. 

On the other hand, when the path loss has a nonzero mean and $K = 1$, recent work \cite{chanisit2024} demonstrates the necessity of having one additional sensing beamformer in some cases. In this situation, removing the extra sensing beamformer would result in a performance loss, thus Assumption~\ref{main_assume} does not hold.  


\subsection{Existing Method and Proposed Solution Strategy}\label{sec:proposed}

As can be seen in the above example, 
the BCRB optimization problem is quite complex.
The goal of this paper is to identify the structure of the 
BCRB optimization problem \eqref{prob:generalCase} that would point toward
an efficient solution. 
In prior work, an SDR strategy for the above AoA estimation example (with the caveat that the classical CRB is used in lieu of the BCRB) has already been studied \cite{LiuFCRB2022}. 
This SDR technique is also applicable to the general problem~\eqref{prob:generalCase} (as will be shown in the next section). 
However, solving the problem in terms of SDR can be rather inefficient. This is because it lifts the optimization space from ${N}_\text{T}K$ complex dimensions to $N_\text{T}^2K$ complex dimensions by defining the PSD matrices $\mathbf{R}_k = \mathbf{v}_k \mathbf{v}_k^\mathsf{H}$. As a result, solving the SDR using interior-point methods would require 
$\mathcal{O}(N_\text{T}^{6} K^3)$ operations per iteration~\cite{boyd2004convex}. This order of complexity is often infeasible in practical implementations, especially when $N_\text{T}$ is large.

In this paper, we take the SDR as a starting point and propose a further
reformulation of problem~\eqref{prob:generalCase} that lends itself 
towards an efficient solution. The proposed solution strategy comprises
two crucial ingredients: (i) a reformulation of the BCRB minimization
objective as a weighted power maximization problem in sensing directions
of interest; and (ii) a connection with the following classical problem 
in multiuser downlink MIMO communications with no sensing
\begin{subequations}\label{prob:dl_comm}
    \begin{align}
            \underset{\mathbf{V}}{\mathrm{minimize}} ~~~~~& \sum_k 
{\mathbf{v}}_k^H \mathbf{\tilde Q} {\mathbf{v}}_k 
\label{eq:comm_obj} \\
            \mathrm{subject \ to}  ~~~~ & \frac{\left|\mathbf{h}_k^\textsf{H} \mathbf{v}_k \right|^2}{\sum_{i \neq k} \left|\mathbf{h}_k^\textsf{H} \mathbf{v}_i \right|^2 + \sigma_\text{c}^2} \geq \gamma_k, \quad \forall k
    \end{align}
\end{subequations}
where the goal is to design downlink beamformers to satisfy SINR constraints
while minimizing a weighted transmit power at the BS with $\mathbf{\tilde Q} \succcurlyeq 0$ as the weighting matrix. 

For minimizing the sum power (i.e., when $\mathbf{\tilde Q} = \mathbf{I}$), a duality notion has been developed in \cite{rashidDL1998} that allows the downlink problem~\eqref{prob:dl_comm} to be transformed to a virtual uplink problem, a result now commonly referred to as uplink-downlink duality~\cite{rashidDL1998, Schubertsolution2004, Viswanathsum2003, Vishwanathduality2003, yutransmitter2007}. 
The key advantage of such a transformation is that the uplink problem is more
computationally efficient to solve than its downlink counterpart.  The
duality relationship for arbitrary $\mathbf{\tilde Q} \succcurlyeq 0$ is developed in \cite{yutransmitter2007}.  This paper further develops the uplink-downlink duality theory for the BCRB minimization problem, for which the effective $\mathbf{\tilde Q}$ may not necessarily be PSD. We show how the sensing objective in \eqref{prob:generalCase} can
be transformed, and how the classical uplink-downlink duality theory can be
extended for the transformed problem. 

\section{Minimization of BCRB as Downlink Power Maximization} 
\label{sec:pw_max}

We begin by deriving a key result that under Assumption~\ref{main_assume}, the minimization of the BCRB in the ISAC problem can be reformulated as a power maximization problem in the sensing directions of interest. 

\subsection{Semidefinite Relaxation}

The first step is to relax the optimization problem (\ref{prob:generalCase}) using SDR. Define $\mathbf{R}_k = \mathbf{v}_k\mathbf{v}_k^\textsf{H}$.  The relaxation of 
the problem (\ref{prob:generalCase}) together with the derivation of the BCRB in (\ref{eq:FIM_elements})-(\ref{eq:d_elements}) is
\begin{subequations}\label{prob:general_SDP_main_text}
    \begin{align}
\underset{\mathbf{R}_1,\ldots, \mathbf{R}_K}{\mathrm{minimize}} ~~&  \Tr \left(\mathbf{W} \mathbf{J}_\mathbf{R}^{-1} \right) 
    \label{eq:sdp_obj} \\
    \mathrm{subject \ to}   ~& \frac{1}{\gamma_k}  \mathbf{h}_k^\textsf{H} \mathbf{R}_k \mathbf{h}_k - \sum_{i \neq k}  \mathbf{h}_k^\textsf{H} \mathbf{R}_i \mathbf{h}_k \geq \sigma_\text{c}^2 \label{eq:sdp_const_1} \\
& \sum_k \Tr \left( \mathbf{R}_k \right) \leq P, \quad \mathbf{R}_k \succcurlyeq 0 ~~\forall k, \label{eq:sdp_const_2}
    \end{align}
\end{subequations}
where $\mathbf{J}_\mathbf{R}$ denotes, with a slight abuse of notation, the BFIM matrix expressed in $\mathbf{R}_1, \ldots, \mathbf{R}_k$ (i.e., with $\mathbf{V}\mathbf{V}^\textsf{H}$ in~\eqref{eq:d_elements} replaced by $\sum_k \mathbf{R}_k$), and the rank-one constraints on $\mathbf{R}_k$ are dropped. Note that $\tfrac{1}{\gamma_k}$ is finite, since $\gamma_k > 0$ for all $k$, so the problem is well defined.

For general quadratic optimization problems, the SDRs do not necessarily provide a tight relaxation of the original problem, because a rank-one solution of the SDR is not guaranteed to exist. However, under Assumption~\ref{main_assume}, the SDR must be tight. We state this result as a lemma. 

\begin{lemma}
\label{lem:SDR_tight}
Under Assumption~\ref{main_assume}, the ISAC problem~\eqref{prob:generalCase} is equivalent to its SDR~\eqref{prob:general_SDP_main_text}. In other words, the SDR~\eqref{prob:general_SDP_main_text} always has a rank-one solution.
    \begin{IEEEproof}
     See Appendix~\ref{Appen:A}. 
\end{IEEEproof}
\end{lemma}


The above result allows us to work with the optimization over the covariance matrices
$\{\mathbf{R}_i\}$ instead of the beamforming vectors $\{\mathbf{V}_i\}$. 
The main advantage here is that the SDR \eqref{prob:general_SDP_main_text} is convex.
This is because
$\mathbf{J}_\mathbf{R}$ is an affine function of $\mathbf{R}_k$ by \eqref{eq:FIM_elements} and the trace-inverse function is convex, so the objective function is convex in  $\mathbf{R}_k$; also, the constraints are linear in $\mathbf{R}_k$. Furthermore, assuming that
the SINR constraints are strictly feasible, so Slater's condition is satisfied, 
the problem \eqref{prob:general_SDP_main_text} in fact has strong duality.

Next, we reformulate the objective function in \eqref{prob:general_SDP_main_text}. 
A useful technique to deal with the trace of a matrix inverse is to use the following
Schur complement relation \cite{vandenberghe1998determinant}
\begin{equation}
\left[ \begin{matrix} \mathbf{J}_{\mathbf{R}} & \boldsymbol{e}_\ell \\ \boldsymbol{e}_\ell^{T} & d_\ell \end{matrix} \right] \succcurlyeq 0 \ \ \Leftrightarrow \ \ d_\ell \geq 0, ~ d_\ell - \mathbf{e}_\ell^{T} \mathbf{J}_{\mathbf{R}}^{-1} \mathbf{e}_\ell \geq 0,
\end{equation}
which allows us to express the objective function \eqref{eq:sdp_obj} as the following minimization problem 
\begin{subequations}\label{prob:trace_inv}
\begin{align}
     \Tr \left(\mathbf{W} \mathbf{J}_\mathbf{R}^{-1} \right) = \min_{d_1, \ldots, d_L} ~~ &  \sum_\ell d_\ell, \\
     \mathrm{s.t.~ }  ~~~~ & \left[ \begin{matrix} \mathbf{J}_\mathbf{R} & \sqrt{w}_\ell e_\ell \\ \sqrt{w}_\ell e_\ell^\mathsf{T} & d_\ell \end{matrix} \right] \succcurlyeq 0, ~\forall \ell.  \label{eq:SDP_const_main_text}
\end{align}
\end{subequations}
This technique is first used in the earlier work on using CRB to optimize waveform for MIMO radar \cite{LiR2008ange} and has since been applied extensively in radar and ISAC settings \cite{Huleihel2013optimal, kokke2023, LiuFCRB2022}.

Now, substituting~\eqref{prob:trace_inv} into~\eqref{prob:general_SDP_main_text}, we obtain
\begin{subequations}\label{prob:general_SDP2_main_text}
    \begin{align}
    \underset{ \mathbf{R}_1,\ldots, \mathbf{R}_K, d_1, \ldots, d_L}{\mathrm{minimize}} ~~ &  \sum_\ell d_\ell  \\
    \mathrm{subject \ to} ~~~~~ & \eqref{eq:sdp_const_1}, \eqref{eq:sdp_const_2}, \eqref{eq:SDP_const_main_text}.
    \end{align}
\end{subequations} 
Problem \eqref{prob:generalCase} is now reformulated into a form that can be solved by a standard semidefinite programming solver. When Assumption~\ref{main_assume} holds, this SDR must have a rank-one solution.
However, as already mentioned, SDR has high complexity. 
In this paper, we take this SDR formulation as the starting point, and proceed by analyzing its structure by taking its Lagrangian dual. This would eventually enable algorithmic development in the beamformer domain rather than the covariance domain.

\subsection{Analysis of the Dual Problem}

The reformulated problem~\eqref{prob:general_SDP2_main_text} is convex with strong duality when $\gamma_k > 0$ and the constraints are strictly feasible. We proceed by taking its Lagrangian dual with respect to \eqref{eq:SDP_const_main_text}. 
Let $\tilde{\mathbf{B}}_1, \ldots, \tilde{\mathbf{B}}_L \in \mathbb{R}^{(L + 1) \times (L + 1)}$ denote the dual variables associated with~\eqref{eq:SDP_const_main_text} 
\begin{equation}
    \tilde{\mathbf{B}}_\ell \triangleq \left[\begin{matrix}
        \mathbf{B}_\ell & -\boldsymbol{\beta}_\ell \\
        -\boldsymbol{\beta}_\ell^\mathsf{T} & b_\ell 
    \end{matrix} \right]   \succcurlyeq 0,
\end{equation}
with $\mathbf{B}_\ell \in \mathbb{R}^{L \times L}$, $ \boldsymbol{\beta}_\ell \in \mathbb{R}^{L}$, and $b_\ell \in \mathbb{R}$. 

The Lagrangian dual of the problem \eqref{prob:general_SDP2_main_text} 
can then be written down as
\begin{align*}
\underset{\tilde{\mathbf{B}}_1, \ldots, \tilde{\mathbf{B}}_L}{\mathrm{maximize}}\min_{\underset{\{d_\ell\} }{ \{\mathbf{R}_k\} \in \mathcal{R},}}   \sum_\ell d_\ell (1 - b_\ell) +2\sqrt{w_\ell} \mathbf{e}_\ell^\mathsf{T} \boldsymbol{\beta}_\ell 
     -\Tr(\mathbf{B}_\ell\mathbf{J}_\mathbf{R})
    \end{align*}
where $\mathcal{R}$ denotes the constraints~\eqref{eq:sdp_const_1}-\eqref{eq:sdp_const_2}. 

Consider the inner minimization of $d_\ell$ for fixed $b_\ell$, then the outer maximization over $b_\ell$. It is easy to conclude that the optimal 
$b^*_\ell = 1, \forall \ell$, as otherwise the inner problem is unbounded below. 
Substituting $b_\ell^* = 1$, we get
\begin{align}\label{eq:max_min_R}
\underset{\mathbf{B}_\ell \succcurlyeq \boldsymbol{\beta}_\ell \boldsymbol{\beta}_\ell^\mathsf{T}, \forall \ell}{\mathrm{maximize}}~~\min_{ \{\mathbf{R}_k\} \in \mathcal{R} }    \sum_\ell  2 \sqrt{w_\ell} \mathbf{e}_\ell^\mathsf{T} \boldsymbol{\beta}_\ell 
     -\Tr(\mathbf{B}_\ell\mathbf{J}_\mathbf{R}),
    \end{align}
 where the condition $\mathbf{B}_\ell \succcurlyeq \boldsymbol{\beta}_\ell \boldsymbol{\beta}_\ell^\mathsf{T}$ follows from
    $\tilde{\mathbf{B}}_\ell \succcurlyeq 0$ and $b_\ell = 1$, so the Schur complement must satisfy $\mathbf{B}_\ell - \boldsymbol{\beta}_\ell \boldsymbol{\beta}_\ell^\mathsf{T} \succcurlyeq 0$. 

Now, since strong duality holds, the primal optimal $\mathbf{R}^*_1, \ldots, \mathbf{R}^*_K$ and the dual optimal $\mathbf{B}_1^*, \ldots, \mathbf{B}^*_L$, $\boldsymbol{\beta}^*_1, \ldots, \boldsymbol{\beta}^*_L$ constitute a saddle-point solution of the max-min problem. Strong duality implies that we can interchange $\min$ and $\max$, i.e.,
	(\ref{eq:max_min_R}) is equivalent to
\begin{align}
	\label{eq:min_max_R}
	\min_{ \{\mathbf{R}_k\} \in \mathcal{R} }    ~~
\underset{\mathbf{B}_\ell \succcurlyeq \boldsymbol{\beta}_\ell \boldsymbol{\beta}_\ell^\mathsf{T}, \forall \ell}{\mathrm{maximize}} ~~ 
	\sum_\ell  2 \sqrt{w_\ell} \mathbf{e}_\ell^\mathsf{T} \boldsymbol{\beta}_\ell 
     -\Tr(\mathbf{B}_\ell\mathbf{J}_\mathbf{R}).
    \end{align}
In particular, for any fixed $\mathbf{R}_1, \ldots, \mathbf{R}_K$, we 
can optimize over $\mathbf{B}_\ell$ by noticing that 
\begin{equation}
        \mathbf{B}_\ell \succcurlyeq \boldsymbol{\beta}_\ell \boldsymbol{\beta}_\ell^\mathsf{T} \ \ \Rightarrow \ \ \Tr(\mathbf{B}_\ell \mathbf{J}_\mathbf{R})   \geq \Tr(\boldsymbol{\beta}_\ell \boldsymbol{\beta}_\ell^\mathsf{T} \mathbf{J}_\mathbf{R}),
\end{equation}
which holds because $\mathbf{J}_\mathbf{R}$ is positive definite.  Thus, the optimal $\mathbf{B}_\ell$ must be $\mathbf{B}_\ell = \boldsymbol{\beta}_\ell \boldsymbol{\beta}^\mathsf{T}_\ell$ at the saddle-point. Substituting this relation into \eqref{eq:min_max_R}, and again interchanging maximization and minimization, we arrive at
\begin{align}\label{prob:general_SDP4_main_text}
\underset{\boldsymbol{\beta} \in \mathbb{R}^{L \times L}}{\mathrm{maximize}}~~\min_{\{\mathbf{R}_k\} \in \mathcal{R}}   ~~ \sum_\ell \left( 2\sqrt{w_\ell} \mathbf{e}_\ell^\mathsf{T} \boldsymbol{\beta}_\ell 
     -\  \boldsymbol{\beta}_\ell^\mathsf{T} \mathbf{J}_\mathbf{R} \boldsymbol{\beta}_\ell \right),
\end{align}
where $\boldsymbol{\beta} \triangleq [\boldsymbol{\beta}_1, \ldots, \boldsymbol{\beta}_L]\in \mathbb{R}^{L \times L}$. 
This max-min problem is equivalent to the SDR problem \eqref{prob:general_SDP_main_text}),
and under Assumption~\ref{main_assume}, the original ISAC problem~\eqref{prob:generalCase}. 

Furthermore, by the following chain of inequalities, the max-min problem~\eqref{prob:general_SDP4_main_text} can be restated in terms of the beamformers.
Let $p^*$ denote the optimal value of problem~\eqref{prob:generalCase}. 
\begin{subequations}
\begin{align}
    p^* &= \max_{\boldsymbol{\beta} \in \mathbb{R}^{L \times L}} \min_{\{\mathbf{R}_k\} \in \mathcal{R}}  \sum_\ell \left( 2\sqrt{w_\ell} e_\ell^\mathsf{T} \boldsymbol{\beta}_\ell 
     -\  \boldsymbol{\beta}_\ell^\mathsf{T} \mathbf{J}_\mathbf{R} \boldsymbol{\beta}_\ell \right) \label{eq:restate_as_BF_1} \\
     &\leq \max_{\boldsymbol{\beta} \in \mathbb{R}^{L \times L}} \min_{\mathbf{V} \in \mathcal{V}}   \sum_\ell \left( 2\sqrt{w_\ell} e_\ell^\mathsf{T} \boldsymbol{\beta}_\ell 
     -\  \boldsymbol{\beta}_\ell^\mathsf{T} \mathbf{J}_\mathbf{V} \boldsymbol{\beta}_\ell \right) \label{eq:restate_as_BF_2} \\
      &\leq \min_{\mathbf{V} \in \mathcal{V}} \max_{\boldsymbol{\beta} \in \mathbb{R}^{L \times L}}  \sum_\ell \left( 2\sqrt{w_\ell} e_\ell^\mathsf{T} \boldsymbol{\beta}_\ell 
     -\  \boldsymbol{\beta}_\ell^\mathsf{T} \mathbf{J}_\mathbf{V} \boldsymbol{\beta}_\ell \right) \label{eq:restate_as_BF_3} \\
     &= \min_{\mathbf{V} \in \mathcal{V}} \sum_\ell w_\ell e_\ell^\mathsf{T} \mathbf{J}_\mathbf{V}^{-1} e_\ell^\mathsf{T} = p^*, \label{eq:restate_as_BF_4}
\end{align}
\end{subequations}
where \eqref{eq:restate_as_BF_2} follows from restricting the optimization over $\mathbf{R}_k$ to rank-one matrices, \eqref{eq:restate_as_BF_3} follows from the minimax inequality, and \eqref{eq:restate_as_BF_4} follows from maximizing the objective in the right-hand side over $\boldsymbol{\beta}$ for fixed $\mathbf{V}$, where the maximum is attained at 
\begin{equation} \label{eqn:beta_w}
\boldsymbol{\beta}_{\ell} =  \sqrt{w}_\ell \mathbf{J}_\mathbf{V}^{-1} \mathbf{e}_\ell, \quad \forall \ell.
\end{equation} 
Clearly, each of these inequalities has to be an equality. This establishes 
the following main result of this section. 



\begin{theorem}
\label{thm:max_min} 
Consider the optimization problem~\eqref{prob:generalCase}.
Suppose that: 
i) $\gamma_k > 0$ for all $k$; 
ii) the constraints \eqref{eq:SINR_const}-\eqref{eq:pw_const} are strictly feasible;
and iii) Assumption~\ref{main_assume} holds. 
Then, $\mathbf{V}^*$ is a global optimum solution of \eqref{prob:generalCase} if and only if there exists $\boldsymbol{\beta}^* \in \mathbb{R}^{L \times L}$ such that $(\mathbf{V}^*, \boldsymbol{\beta}^*)$ is a saddle-point solution of
    \begin{align} \label{prob:max_min_1}
\underset{\boldsymbol{\beta} \in \mathbb{R}^{L \times L}}{\mathrm{maximize}}~~\min_{\mathbf{V} \in \mathcal{V} } ~~ \sum_{\ell = 1}^L 2\sqrt{w_\ell}  \boldsymbol{\beta}_\ell^\mathsf{T} \mathbf{e}_\ell - \boldsymbol{\beta}_\ell^\mathsf{T} \mathbf{J}_{\mathbf{V}} \boldsymbol{\beta}_\ell
\end{align}
where $\boldsymbol{\beta} \triangleq [\boldsymbol{\beta}_1, \ldots, \boldsymbol{\beta}_L]\in \mathbb{R}^{L \times L}$ and $\mathcal{V}$ denotes the set of feasible beamformers under constraints~\eqref{eq:SINR_const}-\eqref{eq:pw_const},

Furthermore, problem~\eqref{prob:max_min_1} is equivalent to
   \begin{align} \label{prob:max_min}
    \underset{\boldsymbol{\beta} \in \mathbb{R}^{L \times L}}{\mathrm{maximize}}~\min_{\mathbf{V} \in \mathcal{V}}\sum_{\ell = 1}^L 2\sqrt{w_\ell}   \boldsymbol{\beta}_\ell^\mathsf{T} \mathbf{e}_\ell -\boldsymbol{\beta}_{\ell}^\mathsf{T} \mathbf{C} \boldsymbol{\beta}_{\ell} - \Tr{\left(\mathbf{Q}_{\boldsymbol{\beta}} \mathbf{V}\mathbf{V}^H \right)}
\end{align}
 where $\mathbf{C}$ is defined in~\eqref{eq:c_elements} and 
\begin{equation}
\mathbf{Q}_{\boldsymbol{\beta}} \triangleq \frac{2T}{\sigma_\text{s}^2}  \sum_{\ell = 1}^L  \mathbb{E}_{\boldsymbol{\eta}} \left[{(\tilde{\mathbf{G}}_{\boldsymbol{\beta_\ell}}^{\boldsymbol{(\eta)}})}^\textsf{H}\tilde{\mathbf{G}}_{\boldsymbol{\beta_\ell}}^{(\boldsymbol{\eta})}\right],
\end{equation}
and 
\begin{equation} 
\tilde{\mathbf{G}}_{\boldsymbol{\beta_\ell}}^{(\boldsymbol{\eta})}\triangleq \sum_i [\boldsymbol{\beta}_\ell]_i \dot{\mathbf{G}}^{(\boldsymbol{\eta})}_i.
\end{equation} 

\begin{figure}
    \centering    \includegraphics[width=0.4\textwidth]{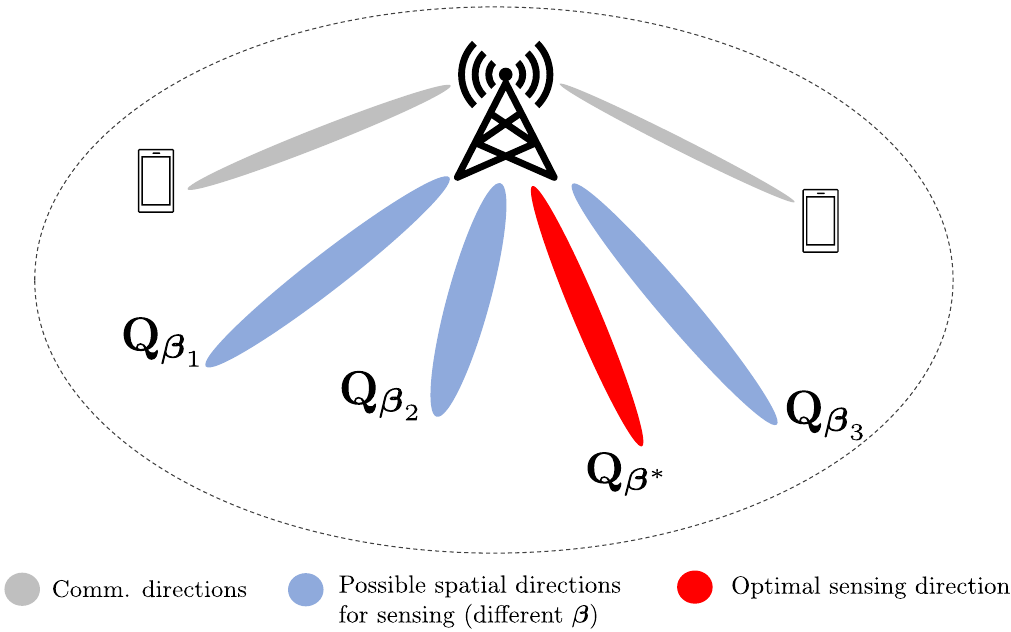}
     \caption{An interpretation of BCRB minimization for the ISAC problem. Different $\boldsymbol{\beta}$ values correspond to different spatial directions. There exists some $\boldsymbol{\beta}^*$ for which the minimization of BCRB can be viewed as maximization of power along a spatial direction of interest, subject to the SINR constraints for the communication users and the total power constraint.}
     \label{fig:interpretation_BCRB}
\end{figure}

\begin{IEEEproof} 
%
The development leading to the theorem shows that \eqref{prob:max_min_1} is the Lagrangian
dual of \eqref{prob:generalCase} under the stated assumptions. Thus, $\mathbf{V}^*$ is an optimal solution of \eqref{prob:generalCase} if and only if there exists $\boldsymbol{\beta}^* \in \mathbb{R}^{L \times L}$ such that $(\mathbf{V}^*, \boldsymbol{\beta}^*)$ is a saddle-point of \eqref{prob:max_min_1}.
To further expand the term involving $\mathbf{J}_{\mathbf{V}}$, we note from 
\eqref{eq:FIM_elements} that
\begin{equation}\label{eq:beta_relation}
    \sum_{\ell = 1}^L \boldsymbol{\beta}_\ell^\mathsf{T} \mathbf{J}_\mathbf{V} \boldsymbol{\beta}_\ell =    \sum_{\ell = 1}^L \boldsymbol{\beta}_\ell^\mathsf{T} \mathbf{C} \boldsymbol{\beta}_\ell +    \sum_{\ell = 1}^L \boldsymbol{\beta}_\ell^\mathsf{T} \mathbf{T}_\mathbf{V} \boldsymbol{\beta}_\ell,
\end{equation}
and
\begin{subequations}\label{eq:D-Qrelation}
\begin{align}
    &\sum_{\ell = 1}^L  \boldsymbol{\beta}_\ell^\mathsf{T} \mathbf{T}_\mathbf{V} \boldsymbol{\beta}_\ell 
= \sum_{\ell = 1}^L  \sum_{ij} [\boldsymbol{\beta}_\ell]_i  \left[\mathbf{T}_\mathbf{V}\right]_{ij}  [\boldsymbol{\beta}_\ell]_j \\
    &= \frac{2T}{\sigma_\text{s}^2} \Tr\left( \sum_{\ell = 1}^L   \sum_{ij} [\boldsymbol{\beta}_\ell]_i  \mathbb{E}_{\boldsymbol{\eta}}\left[({\dot{\mathbf{G}}^{(\boldsymbol{\eta})}_i})^\textsf{H} \dot{\mathbf{G}}^{(\boldsymbol{\eta})}_j\right] \mathbf{V} \mathbf{V}^\textsf{H} [\boldsymbol{\beta}_\ell]_j  \right) \label{eq:2c} \\
    &= \frac{2T}{\sigma_\text{s}^2}\Tr\left(  \sum_{\ell = 1}^L  \mathbb{E}_{\boldsymbol{\eta}}\left[ \sum_{i} [\boldsymbol{\beta}_\ell]_i ({\dot{\mathbf{G}}^{(\boldsymbol{\eta})}_i})^\textsf{H}  \sum_j [\boldsymbol{\beta}_\ell]_j  \dot{\mathbf{G}}^{(\boldsymbol{\eta})}_j \right] \mathbf{V} \mathbf{V}^\textsf{H}  \right) \\
    &= \frac{2T}{\sigma_\text{s}^2}\Tr\left(\sum_{\ell = 1}^L   \mathbb{E}_{\boldsymbol{\eta}}\left[{(\tilde{\mathbf{G}}_{\boldsymbol{\beta_\ell}}^{\boldsymbol{(\eta)}})}^\textsf{H}  \tilde{\mathbf{G}}_{\boldsymbol{\beta_\ell}}^{(\boldsymbol{\eta})}\right] \mathbf{V} \mathbf{V}^\textsf{H}  \right) \\
    &= \Tr\left(  \mathbf{Q}_{\boldsymbol{\beta}} \mathbf{V} \mathbf{V}^\textsf{H}  \right) \label{eq:2d}
\end{align}
\end{subequations}
where \eqref{eq:2c} follows from~\eqref{eq:d_elements}. 
\end{IEEEproof}
\end{theorem}

Theorem~\ref{thm:max_min} allows us to reformulate the ISAC problem~\eqref{prob:generalCase} in a more tractable form as the saddle-point solution of a quadratic program as expressed in \eqref{prob:max_min}, under Assumption~\ref{main_assume}. 
Furthermore, Theorem~\ref{thm:max_min} offers the following important interpretation of BCRB optimization. 

Examining the max-min problem~\eqref{prob:max_min}, by construction the matrix $\mathbf{Q}_{\boldsymbol{\beta}}$ is PSD for any $\boldsymbol{\beta}$. Consequently, the trace term can be interpreted as a sum of beamforming powers measured with respect to the matrix $\mathbf{Q}_{\boldsymbol{\beta}}$. Thus, for a fixed $\boldsymbol{\beta}$, the inner problem in~\eqref{prob:max_min} is a power maximization
along certain directions of interest. 
Varying $\boldsymbol{\beta}$ corresponds to varying the directions of interest (as illustrated in Fig.~\ref{fig:interpretation_BCRB}). Theorem~\ref{thm:max_min} states that 
there exists a judicious choice $\boldsymbol{\beta}^*$ for which the problem of minimizing the BCRB has the same solution as that of maximizing weighted power in the direction of  $\mathbf{Q}_{\boldsymbol \beta^*}$.




\subsection{AoA Estimation Example}

Consider again the problem of estimating the AoA for a single target as given in problem~\eqref{prob:simpleCase}. For this example, recall that Assumption~\ref{main_assume} holds whenever $K \ge 2$. 
In this case, Theorem~\ref{thm:max_min} can be applied to reformulate problem~\eqref{prob:simpleCase} as a max-min problem.

Since $L = 3$, we have $\boldsymbol{\beta} = \left[\boldsymbol{\beta}_1, \boldsymbol{\beta}_2, \boldsymbol{\beta}_3 \right]$, where $\boldsymbol{\beta}_\ell \in \mathbb{R}^{3}$. 
We can further set $\boldsymbol{\beta}_1 
 = \boldsymbol{\beta}_2 = \boldsymbol{0}$ in this case, because the preceding analysis \eqref{eqn:beta_w} shows that the optimal $\boldsymbol{\beta}_\ell$ is given by $\boldsymbol{\beta}_\ell^* = \sqrt{w_\ell} \mathbf{J}_{\mathbf{V}*}^{-1} \mathbf{e}_\ell$, and we have $w_1 = w_2 = 0$, since we are interested in estimating $\theta$ only and not $\alpha$.  
Then, the set of auxiliary variables reduces to $\boldsymbol{\beta}_3 = [b_1, b_2, b_3]^\mathsf{T}$ only. By applying Theorem~\ref{thm:max_min}, problem~\eqref{prob:simpleCase} can be equivalently expressed as
  \begin{align} \label{prob:max_min_special_case}
    \underset{\boldsymbol{\beta}_3 \in \mathbb{R}^{3}}{\mathrm{maximize}}~\min_{\mathbf{V} \in \mathcal{V}} ~ 2   b_3 -\boldsymbol{\beta}_3^\mathsf{T} \mathbf{C} \boldsymbol{\beta}_3 - \Tr{\left(\mathbf{Q}_{\boldsymbol{\beta}_3} \mathbf{V}\mathbf{V}^H \right)}
\end{align}
where 
\begin{equation}
\mathbf{Q}_{\boldsymbol{\beta}_3} = \frac{2T}{\sigma_\text{s}^2} \mathbb{E}_{\alpha, \theta}\left[ \left(\tilde{\mathbf{G}}^{(\alpha,\theta)}_{\boldsymbol{\beta}_3}\right)^\textsf{H} \tilde{\mathbf{G}}^{(\alpha,\theta)}_{\boldsymbol{\beta}_3} \right]
\end{equation}
and
\begin{equation}
\tilde{\mathbf{G}}^{(\alpha,\theta)}_{\boldsymbol{\beta}} =  \left( (b_1 + \imath b_2) \mathbf{A}(\theta)  + b_3 \alpha \dot{\mathbf{A}}(\theta) \right). 
\end{equation}  
We further transform the problem by interchanging max and min,
then introducing a new variable $b = \tfrac{(b_1 + \imath b_2)}{b_3}$,
which turns out to allow a simplification of problem \eqref{prob:max_min_special_case} as 
\begin{align} \label{prob:max_min_AoA_final_form} 
    \underset{b\in \mathbb{C}}{\mathrm{minimize}}~~\max_{\mathbf{V} \in \mathcal{V}} ~~ \Tr\left( \mathbf{Q}_b \mathbf{V} \mathbf{V}^\textsf{H} \right) 
    + \frac{|b|^2}{\sigma_\alpha^2},
\end{align}
where $b \in \mathbb{C}$ is the new variable and
\begin{equation}
\label{prob:max_min_AoA_final_form_Qb} 
\mathbf{Q}_b \triangleq \mathbb{E}_{\alpha, \theta}\left[ ( b \mathbf{A}(\theta) + \alpha \dot{\mathbf{A}}(\theta) )^\textsf{H} ( b \mathbf{A}(\theta) + \alpha \dot{\mathbf{A}}(\theta) )  \right]. 
\end{equation}
The equivalence between problem~\eqref{prob:max_min_special_case} and problem~\eqref{prob:max_min_AoA_final_form} is established in Appendix~\ref{appen:simple_minimax_formulation_AoA} by utilizing the fact that $\mathbf{C} \succ  0$, which holds whenever the priors have finite variance. 


The above formulations illustrate that the beamforming strategy that minimizes the BCRB also maximizes the power in the direction of some PSD matrix (i.e., the matrix $\mathbf{Q}_{\boldsymbol{\beta}_3} \succcurlyeq 0$ in case of problem~\eqref{prob:max_min_special_case}, or the matrix 
$\mathbf{Q}_b \succcurlyeq 0$ in case of problem~\eqref{prob:max_min_AoA_final_form}). This observation is consistent with the interpretation of Theorem~\ref{thm:max_min}. 
Observe that even though we wish to estimate $\theta$ only, the optimal beamforming strategy depends on the distribution of $\alpha$ through the matrices $\mathbf{Q}_{\boldsymbol{\beta}_3}$ and $\mathbf{Q}_b$. 

The expression of $\mathbf{Q}_b$ also sheds light on a certain heuristic used in beamforming design for AoA estimation, e.g.,~\cite{soticaprobing2007, wen2023}. In particular, previous works propose to optimize the signal power, i.e., $\Tr{( \mathbb{E}_{\theta} [ \mathbf{A}(\theta) \mathbf{A}^\textsf{H}(\theta)]  \mathbf{V} \mathbf{V}^\textsf{H} )}$, as a heuristic metric to guide the beamforming design. The preceding expression reveals that this heuristic choice, while reasonable, is also not necessarily optimal. 
Specifically, it shows that optimal direction depends not only on the direction of maximal power as given by the array response $\mathbf{A}(\theta)$, but also on the \emph{derivative} of $\mathbf{A}(\theta)$ with respect to the parameter of interest $\theta$. Indeed, the optimal direction should depend on some weighted combination of maximizing the \emph{signal power} in the expected AoA and its \emph{sensitivity} with respect to the AoA, with weights controlled by the outer minimization in \eqref{prob:max_min_AoA_final_form}.



\section{Uplink-Downlink Duality for ISAC}\label{sec:duality}

We now proceed to establish an uplink-downlink duality relation for the
equivalent formulation \eqref{prob:max_min} of the ISAC problem.  This is
to say that \eqref{prob:max_min} admits an alternate formulation in terms
of a virtual uplink problem in a manner similar to the transformation of
the classical downlink communications problem~\eqref{prob:dl_comm} to an uplink problem
\cite{rashidDL1998, yutransmitter2007}.  However, the duality result for
the ISAC problem also differs significantly from the conventional
uplink-downlink duality for the communications problem in some
important technical aspects as explained below.
\subsection{ISAC as Weighted Downlink Power Minimization} 

We begin by analyzing the inner minimization problem in \eqref{prob:max_min} 
under the SINR and the total power constraints, for fixed $\boldsymbol \beta$, as rewritten explicitly below:
\begin{subequations} \label{prob:inner_min}
\begin{align} \label{prob:inner_min_obj}
\underset{\mathbf{v}_1, \ldots, \mathbf{v}_K}{\mathrm{minimize}} ~~~\; &  
\sum_{\ell = 1}^L 2\sqrt{w_\ell}   \boldsymbol{\beta}_\ell^\mathsf{T} \mathbf{e}_\ell -\boldsymbol{\beta}_{\ell}^\mathsf{T} \mathbf{C} \boldsymbol{\beta}_{\ell} - \Tr{\left(\mathbf{Q}_{\boldsymbol{\beta}} \mathbf{V}\mathbf{V}^H \right)} \\
\mathrm{subject\ to}  ~~~ & \frac{\left|\mathbf{h}_k^\textsf{H} \mathbf{v}_k \right|^2}{\sum_{i \neq k} \left|\mathbf{h}_k^\textsf{H} \mathbf{v}_i \right|^2 + \sigma_\text{c}^2} \geq \gamma_k, \quad \forall k \\
            & \Tr\left( \mathbf{V} \mathbf{V}^\mathsf{H} \right)  \leq P. 
\end{align}
\end{subequations}
By dualizing with respect to the total power constraint, the Lagrangian of this optimization problem can be written as 
\ifdraftmode
\begin{equation}\label{eq:lagrangian}
\mathcal{L}_{\lambda, \boldsymbol{\beta}}\left( \mathbf{V}\right) = \sum_{k=1}^K \mathbf{v}_k^\textsf{H} \left(\lambda \mathbf{I} - \mathbf{Q}_{\boldsymbol{\beta}}  \right)  \mathbf{v}_k - \lambda P + \sum_{\ell = 1}^L 2 \sqrt{w}_\ell \boldsymbol{\beta}_\ell^\mathsf{T} \mathbf{e}_\ell -\boldsymbol{\beta}_{\ell}^\mathsf{T} \mathbf{C} \boldsymbol{\beta}_{\ell}. 
\end{equation}
\else
\begin{multline}\label{eq:lagrangian}
    \mathcal{L}_{\lambda, \boldsymbol{\beta}}\left( \mathbf{V}\right) = \sum_{k=1}^K \mathbf{v}_k^\textsf{H} \left(\lambda \mathbf{I} - \mathbf{Q}_{\boldsymbol{\beta}}  \right)  \mathbf{v}_k - \lambda P \\ + \sum_{\ell = 1}^L 2 \sqrt{w}_\ell \boldsymbol{\beta}_\ell^\mathsf{T} \mathbf{e}_\ell -\boldsymbol{\beta}_{\ell}^\mathsf{T} \mathbf{C} \boldsymbol{\beta}_{\ell}. 
\end{multline}
\fi
The constrained minimization problem \eqref{prob:inner_min} can be thought of as a minimization
of $\mathcal{L}_{\lambda, \boldsymbol{\beta}}\left( \mathbf{V}\right)$ over $\mathbf V$ together with an
outer maximization over $\lambda$. Now the maximization over $\lambda$ can be combined 
with the outer maximization over $\boldsymbol \beta$ in \eqref{prob:max_min}, so the overall
problem \eqref{prob:max_min} can be equivalently written as
\begin{subequations}\label{prob:partial_dual}
\begin{align} 
\underset{\boldsymbol{\beta}, \lambda \geq 0 }{\mathrm{maximize}}~~~\min_{\mathbf{V}}~~~~~&\mathcal{L}_{\lambda, \boldsymbol{\beta}}\left( \mathbf{V}\right) \\ 
 \mathrm{s.t.\ }  ~~~~~ & 
\frac{\left|\mathbf{h}_k^\textsf{H} \mathbf{v}_k \right|^2}{\sum_{i \neq k} \left|\mathbf{h}_k^\textsf{H} \mathbf{v}_i \right|^2 + \sigma_\text{c}^2} \geq \gamma_k, \quad \forall k. 
\end{align}
\end{subequations}
%
This new problem~\eqref{prob:partial_dual} attains the same optimal value as
the original ISAC problem~\eqref{prob:generalCase}. Furthermore, at the optimal 
$(\lambda^*, \boldsymbol{\beta}^*)$ of \eqref{prob:partial_dual}, the optimal 
beamforming strategy of the ISAC problem \eqref{prob:generalCase} must be 
a minimizer of $\mathcal{L}_{\lambda^*, \boldsymbol{\beta}^*}\left( \cdot\right)$. 
This is because under Assumption~\ref{main_assume}, the ISAC problem 
\eqref{prob:generalCase} has strong duality (as proved by noting 
that its SDR is tight, in Lemma~\ref{lem:SDR_tight}.)


Now, fix $(\lambda, \boldsymbol{\beta})$ and consider the inner minimization problem in~\eqref{prob:partial_dual}. Keeping only the terms involving $\mathbf{v}_k$, we have%
\begin{subequations}\label{prob:fixed_lambda}
\begin{align}
\underset{\mathbf{V}}{\mathrm{minimize}}~~~&  \sum_{k=1}^K \mathbf{v}_k^\textsf{H} \left(\lambda \mathbf{I} - \mathbf{Q}_{\boldsymbol{\beta}}  \right)  \mathbf{v}_k  \label{prob:fixed_lambda_obj} \\ 
 \mathrm{subject\ to\ }  ~~& 
\frac{\left|\mathbf{h}_k^\textsf{H} \mathbf{v}_k \right|^2}{\sum_{i \neq k} \left|\mathbf{h}_k^\textsf{H} \mathbf{v}_i \right|^2 + \sigma_\text{c}^2} \geq \gamma_k, \quad \forall k.
\end{align}
\end{subequations}
This problem is now in the same form as the weighted downlink power minimization problem \eqref{prob:dl_comm} with weight $\mathbf{\tilde Q} =\lambda \mathbf{I} -\mathbf{Q}_{\boldsymbol{\beta}}$, for which an uplink-downlink duality relationship has been established in \cite{yutransmitter2007}, 
\emph{except} the conventional downlink power minimization problem requires $\mathbf{\tilde Q} \succcurlyeq 0$ in \eqref{prob:dl_comm}, but this is not necessarily true for the ISAC problem \eqref{prob:fixed_lambda}. 

The condition that $\mathbf{\tilde Q}$ must be PSD is important in the result of \cite{yutransmitter2007}, because 
the weight matrix $\mathbf{\tilde Q}$ for the downlink power minimization problem becomes a noise covariance matrix in the dual uplink channel.
Without such a condition, the noise in the uplink problem does not have a physical meaning. 

For the ISAC problem, there is no guarantee that for arbitrary $\lambda$ and $\boldsymbol \beta$, we would have $\lambda \mathbf{I} -\mathbf{Q}_{\boldsymbol{\beta}} \succcurlyeq 0$.  
In fact, it is essential to properly deal with the cases where 
$\lambda \mathbf{I} -\mathbf{Q}_{\boldsymbol{\beta}}$ is not PSD for sensing tasks, 
because the sensing
objective typically requires \emph{maximizing} power in certain spatial directions,
rather than \emph{minimizing} power as in traditional downlink problems.
So, $\lambda \mathbf{I} -\mathbf{Q}_{\boldsymbol{\beta}}$ is typically not PSD for the ISAC problem.

Nevertheless, we still need to restrict the space of $(\lambda, \boldsymbol \beta)$ 
under consideration. In this paper, we define the following notion of admissibility, 
and show that the optimal $(\lambda^*, \boldsymbol{\beta}^*)$ is always admissible, and 
moreover, an uplink-downlink duality relationship can be established for all admissible  $(\lambda, \boldsymbol \beta)$'s.
\begin{define}\label{define:admissibility}
The pair $(\lambda, \boldsymbol{\beta})$ is said to be admissible 
for a set of channels $(\mathbf{h}_1, \cdots, \mathbf{h}_K)$
under the SINR thresholds $(\gamma_1, \cdots, \gamma_K)$ for the $K$ users
with noise variance $\sigma_\text{c}^2$, 
if whenever a set of $\mathbf{V} =[\mathbf{v}_1, \ldots, \mathbf{v}_K]$ satisfies 
\begin{equation}\label{eq:sinr_def}
\frac{\left|\mathbf{h}_k^\textsf{H} \mathbf{v}_k \right|^2}{\sum_{i \neq k} \left|\mathbf{h}_k^\textsf{H} \mathbf{v}_i \right|^2 + \sigma_\text{c}^2} \geq \gamma_k, \quad \forall k
\end{equation}
we always have
\begin{equation} \label{eq:quad_sum}
\sum_{k=1}^K
\mathbf{v}_k^\textsf{H} \left( \lambda \mathbf{I} - \mathbf{Q}_{\boldsymbol{\beta}} \right)\mathbf{v}_k  \geq 0, \quad \forall k.
\end{equation}
\end{define}

\editrev{To understand the notion of admissibility, observe that any $(\lambda, \boldsymbol{\beta})$ for which $(\lambda \mathbf{I} - \mathbf{Q}_{\boldsymbol{\beta}})$ is PSD is obviously admissible, since \eqref{eq:quad_sum} is satisfied for any set of beamformers $\mathbf{v}_k$ (including the ones that satisfy the SINR constraints~\eqref{eq:sinr_def}). Likewise, $(\lambda, \boldsymbol{\beta})$ for which  $(\lambda \mathbf{I} - \mathbf{Q}_{\boldsymbol{\beta}})$ is negative definite cannot be admissible.

When $(\lambda, \boldsymbol{\beta})$ is such that $(\lambda \mathbf{I} -
\mathbf{Q}_{\boldsymbol{\beta}})$ is neither positive nor negative definite,
the notion of admissibility is more intricate.  The admissibility condition
(\ref{eq:quad_sum}) suggests that $(\lambda \mathbf{I} -\mathbf{Q}_{\boldsymbol{\beta}})$ 
behaves somewhat like a PSD matrix, but it is a much weaker condition for two reasons.
First, (\ref{eq:quad_sum}) only needs to hold for the set of $\mathbf{v}_k$'s satisfying the SINR constraints (\ref{eq:sinr_def}), and not all $\mathbf{v}_k$'s. Second, only the \emph{sum} power is required to be nonnegative; the individual term $\mathbf{v}_k^\textsf{H} \left( \lambda \mathbf{I} - \mathbf{Q}_{\boldsymbol{\beta}} \right)\mathbf{v}_k$ can actually be negative.  
}


For inadmissible $(\lambda, \boldsymbol{\beta})$, the minimization problem (\ref{prob:fixed_lambda}) would be unbounded below, because one can simply scale a set of $\mathbf{v}_k$'s for which (\ref{eq:quad_sum}) is negative and drive the objective to negative infinity while satisfying the SINR constraints. Clearly, such $(\lambda, \boldsymbol{\beta})$ cannot be a solution to the outer maximization problem in (\ref{prob:partial_dual}), or equivalently, the original problem (\ref{prob:max_min}). 
For this reason, the search over the optimal  $(\lambda, \boldsymbol{\beta})$ in the outer maximization of (\ref{prob:partial_dual}) can be restricted to the admissible set.

More importantly, we can show that whenever $(\lambda, \boldsymbol{\beta})$ is
admissible, an uplink-downlink duality result can be established for 
the inner minimization in (\ref{prob:partial_dual}), or equivalently
(\ref{prob:fixed_lambda}), as explored in detail in the next section.  
Such uplink-downlink duality would have been
a direct consequence of prior work \cite{yutransmitter2007} if 
$\lambda \mathbf{I} - \mathbf{Q}_{\boldsymbol{\beta}} \succcurlyeq 0$, 
in which case $(\lambda \mathbf{I} - \mathbf{Q}_{\boldsymbol{\beta}})$ becomes a
noise covariance in the dual uplink channel. 
The interesting and nontrivial fact is that even if $(\lambda \mathbf{I} - \mathbf{Q}_{\boldsymbol{\beta}})$ is not PSD (so the uplink noise power may be negative in certain directions), as long as $(\lambda, \boldsymbol{\beta})$ is admissible, a duality relationship can still be established.

 It is worth mentioning that the idea of establishing duality for a downlink problem involving power maximization first appeared in~\cite{jie2014} in the context of energy harvesting.  The result developed herein is an extension of that in~\cite{jie2014}. The admissible set in the setting of \cite{jie2014} is in a one-dimensional space. For ISAC problems, the admissible set is in a much higher $1+L\times L$ dimensional space. This makes the optimization over the admissible set in the ISAC setting much more challenging than the energy harvesting problem. 

\subsection{Uplink-Downlink Duality} \label{sec:ul_dl_duality}

The following theorem states an uplink-downlink duality relationship for admissible $(\lambda, \boldsymbol{\beta})$ pairs.
The dual uplink channel here has to satisfy an additional technical condition
related to a certain matrix being an $M$-matrix.
\begin{theorem}
\label{thm:ULDL_duality}
Let $\mathcal{A}$ be the set of admissible $(\lambda, \boldsymbol{\beta})$'s for  
a $K$-user downlink beamforming problem with channels 
$(\mathbf{h}_1, \cdots, \mathbf{h}_K)$, SINR thresholds $(\gamma_1, \cdots, \gamma_K) \in \mathbb{R}_{++}^K$, 
and noise variance $\sigma_\text{c}^2$ 
\begin{subequations}\label{prob:DL}
\begin{align}
\underset{\mathbf{V}}{\mathrm{minimize}}~~~&  \sum_{k=1}^K \mathbf{v}_k^\textsf{H} \left(\lambda \mathbf{I} - \mathbf{Q}_{\boldsymbol{\beta}}  \right)  \mathbf{v}_k  \label{prob:DL_obj} \\ 
 \mathrm{subject\ to\ }  ~~& 
\frac{\left|\mathbf{h}_k^\textsf{H} \mathbf{v}_k \right|^2}{\sum_{i \neq k} \left|\mathbf{h}_k^\textsf{H} \mathbf{v}_i \right|^2 + \sigma_\text{c}^2} \geq \gamma_k, \quad \forall k \label{prob:DL_sinr}
\end{align}
\end{subequations}
where 
the SINR constraints are assumed to be strictly feasible. The optimal value of this downlink beamforming optimization problem
is bounded below if and only if $(\lambda, \boldsymbol{\beta}) \in \mathcal{A}$. 

Furthermore, for $(\lambda, \boldsymbol{\beta}) \in \mathcal{A}$, an optimal solution $\mathbf{V}^*_{\lambda, \boldsymbol{\beta}} \triangleq \left[\mathbf{v}^*_1, \ldots, \mathbf{v}^*_K\right]  \in \mathbb{C}^{N_\text{T} \times K}$  to (\ref{prob:DL}) can be obtained as 
\begin{equation}\label{eq:opt_V_fixed}
\mathbf{v}_k^* = \sqrt{p_k^*} \mathbf{u}_k^*, \quad \forall k,
\end{equation}
where 
$\mathbf{p}^*_{\lambda, \boldsymbol{\beta}} = \left[p^*_1, \ldots, p^*_K \right]$ 
is a set of optimal downlink powers given by 
\begin{equation}
\label{eq:optimal_p}
\mathbf{p}^*_{\lambda, \boldsymbol{\beta}} = \sigma_\text{c}^2 \mathbf{M}_{\mathbf{U}_{\lambda, \boldsymbol{\beta}}^*}^{-1} \mathbf{1},
\end{equation} 
and 
$\mathbf{U}^*_{\lambda, \boldsymbol{\beta}} = \left[ \mathbf{u}_1^*, \ldots, \mathbf{u}_K^*\right] \in \mathbb{C}^{N_\text{T} \times K}$ 
is a set of normalized beamformers that can be obtained from 
any optimal solution to the following uplink problem 
over the beamformers 
$\mathbf{u}_1, \ldots, \mathbf{u}_K$ and uplink transmit powers 
$q_1, \ldots, q_K$: 
\begin{subequations} \label{prob:UL_problem_fixed_lambda}
\begin{align} 
\underset{\mathbf{U}, \; \mathbf{q}}{\mathrm{minimize}}  ~~& \sigma_\text{c}^2  \mathbf{1}^\mathsf{T} \mathbf{q}   \\
    \mathrm{subject \ to} ~& \frac{q_k \left|\mathbf{h}_k^\textsf{H} \mathbf{u}_k \right|^2}{\gamma_k}  \geq \sum_{i \neq k}  q_i \left| \mathbf{h}_i^\textsf{H}\mathbf{u}_k\right|^2 +  \mathbf{u}_k^\textsf{H} 
   (\lambda \mathbf{I} - \mathbf{Q}_{\boldsymbol{\beta}})
    \mathbf{u}_k, 
    \label{eq:ul_sinr} \\
	& \| \mathbf{u}_{k} \| = 1, \quad q_k \ge 0, \quad \forall k \label{eq:normalize} \\
	& \mathbf{M}_{\mathbf U}^{-1} \ge 0. \label{eq:U_M_matrix} 
\end{align}
\end{subequations} 
Here, $\mathbf{M}_\mathbf{U} \in \mathbb{R}^{K \times K}$ denotes a matrix defined as
\begin{equation}
{\left[ \mathbf{M}_\mathbf{U} \right]_{ij} \triangleq} 
\left\{
\begin{array}{cc}
\frac{1}{\gamma_i} \left|\mathbf{h}_i^\textsf{H} \mathbf{u}_i\right|^2, & \mathrm{\ for\quad } i = j, \\ 
- \left| \mathbf{h}_i^\textsf{H} \mathbf{u}_j \right|^2, & \mathrm{\ for\quad } i \neq j,
\end{array}
\right.
\label{eq:M}
\end{equation}
and the notation $\mathbf{M}_\mathbf{U}^{-1} \ge 0$ in constraint~\eqref{eq:U_M_matrix} means that $\mathbf{M}_\mathbf{U}$ is invertible and $\mathbf{M}_\mathbf{U}^{-1}$ has nonnegative entries. 

Finally, the optimal values of the downlink problem \eqref{prob:DL} and the uplink problem \eqref{prob:UL_problem_fixed_lambda} are the same.

\end{theorem}

$\quad$ \emph{Remarks:} The above theorem states that whenever the pair $\left(\lambda, \boldsymbol{\beta}\right)$ is admissible, the downlink solution of problem~\eqref{prob:fixed_lambda} can be recovered by solving problem~\eqref{prob:UL_problem_fixed_lambda}. Before proceeding to the proof, we first justify why problem~\eqref{prob:UL_problem_fixed_lambda} can be referred to as an \emph{uplink problem}, despite that  $\lambda \mathbf{I} - \mathbf{Q}_{\boldsymbol{\beta}}$ is not necessarily PSD. The key reason is that an optimal solution to~\eqref{prob:UL_problem_fixed_lambda} is always such that
\begin{equation}\label{eq:ult_sinr_opt_ineq}
     \sum_{i \neq k}  q^*_i \left| \mathbf{h}_i^\textsf{H}\mathbf{u}^*_k\right|^2 +  {\mathbf{u}^*_k}^\textsf{H} 
   (\lambda \mathbf{I} - \mathbf{Q}_{\boldsymbol{\beta}})
    \mathbf{u}^*_k \geq 0, \quad \forall k.
\end{equation}
In other words, the right-hand side of constraint~\eqref{eq:ul_sinr} is always nonnegative at the optimum. This quantity can be thought of as the combined interference and noise power, hence justifying \eqref{prob:UL_problem_fixed_lambda} as an SINR constraint for the uplink.

Note that~\eqref{eq:ult_sinr_opt_ineq} is trivial to prove if $\lambda \mathbf{I} - \mathbf{Q}_{\boldsymbol{\beta}} \succ 0$, but for the ISAC problem, it is not necessarily true that $\lambda \mathbf{I} - \mathbf{Q}_{\boldsymbol{\beta}} \succ 0$. Below, we show that whenever $(\lambda, \boldsymbol{\beta})$ is admissible, then \eqref{eq:ult_sinr_opt_ineq} always holds. 

The first step is to note that~\eqref{eq:ul_sinr} is always satisfied with equality at the optimum, because if this were not true, then we can always decrease some $q_k$ without violating the constraint, while improving the objective. 
This means that
\begin{equation}\label{eq:ult_sinr_opt_eq}
    \frac{q^*_k}{\gamma_k} \left|\mathbf{h}_k^\textsf{H} \mathbf{u}^*_k \right|^2 = \sum_{i \neq k}  q^*_i \left| \mathbf{h}_i^\textsf{H}\mathbf{u}^*_k\right|^2 +  {\mathbf{u}^*_k}^\textsf{H} 
   (\lambda \mathbf{I} - \mathbf{Q}_{\boldsymbol{\beta}})
    \mathbf{u}^*_k, ~~\forall k.
\end{equation}
Now, using the fact that the optimal $\mathbf{q}^*$ must satisfy $\mathbf{q}^* \geq \mathbf{0}$ as in~\eqref{eq:normalize}, we see that the left-hand side of~\eqref{eq:ult_sinr_opt_eq} is nonnegative, which implies that the right-hand side is also nonnegative. In other words,~\eqref{eq:ult_sinr_opt_ineq} is always satisfied at the optimum as claimed earlier.

We use~\eqref{eq:ult_sinr_opt_ineq} to show that problem~\eqref{prob:UL_problem_fixed_lambda} can be viewed as an uplink problem. Suppose that~\eqref{eq:ult_sinr_opt_ineq} is satisfied with strict inequality for all $k$, i.e., $\sum_{i \neq k}  q^*_i \left| \mathbf{h}_i^\textsf{H}\mathbf{u}_k^*\right|^2 +  {\mathbf{u}^*_k}^\textsf{H} 
   (\lambda \mathbf{I} - \mathbf{Q}_{\boldsymbol{\beta}})
    \mathbf{u}^*_k > 0, \forall k.$
Then, constraint~\eqref{eq:ul_sinr} can be rearranged as 
\begin{subequations}\label{prob:ul_mod}
\begin{align} 
\underset{\mathbf{U}, \; \mathbf{q}}{\mathrm{minimize}}  ~~& \sigma_\text{c}^2  \mathbf{1}^\mathsf{T} \mathbf{q}   \\
    \mathrm{subject \ to} ~& \frac{q_k\left|\mathbf{h}_k^\textsf{H} \mathbf{u}_k \right|^2}{\sum_{i \neq k}  q_i \left| \mathbf{h}_i^\textsf{H}\mathbf{u}_k\right|^2 +  \mathbf{u}_k^\textsf{H} 
   (\lambda \mathbf{I} - \mathbf{Q}_{\boldsymbol{\beta}})\mathbf{u}_k}  \geq \gamma_k,
    \forall k \label{eq:SINR_ul_mod}
    \\
	& ~\eqref{eq:normalize} ~\text{and}~\eqref{eq:U_M_matrix}.
\end{align}
\end{subequations} 
It is now readily seen that the left-hand of \eqref{eq:SINR_ul_mod} is identical to the usual uplink SINR expression with uplink powers $\mathbf{q}_1, \ldots, \mathbf{q}_K$, uplink beamformers $\mathbf{u}_1, \ldots, \mathbf{u}_K$, and noise covariance $\lambda \mathbf{I} - \mathbf{Q}_{\boldsymbol{\beta}}$. It should be noted that $\lambda \mathbf{I} - \mathbf{Q}_{\boldsymbol{\beta}}$ may not be a ``noise covariance" in a standard sense, as it is not necessarily PSD as discussed earlier. Despite this, we loosely use the term to refer to such a matrix.

Based on the above interpretation of constraint~\eqref{eq:SINR_ul_mod}, we can view problem~\eqref{prob:ul_mod} (and hence the equivalent problem~\eqref{prob:UL_problem_fixed_lambda}) as an uplink beamforming and power control problem, similar to that considered in the classical work~\cite{rashidUL1998}. It should be noted, however, that here, there is an additional constraint~\eqref{eq:U_M_matrix}, This constraint is discussed in more detail later in the proof of Theorem~\ref{thm:ULDL_duality}.

When the left-hand side of \eqref{eq:ult_sinr_opt_ineq} is zero for some $k$, we can still view \eqref{prob:UL_problem_fixed_lambda} as an uplink problem. However, if expressed as in \eqref{prob:ul_mod}, the SINR expression for the $k$-th user would have zero in the denominator. 
In fact, in this case the optimal uplink power $q^*_k$ is also zero, so we have zero divided by zero for the SINR expression. This peculiar situation can indeed arise, which is why
we prefer to work with problem~\eqref{prob:UL_problem_fixed_lambda} as opposed to~\eqref{prob:ul_mod}, as the former is always well-defined.

\begin{figure*}
\centering
\subfloat[Downlink beamforming problem for minimizing the transmit power weighted by
$(\lambda \mathbf{I} - \mathbf{Q}_{\boldsymbol{\beta}})$, while satisfying the downlink SINR constraints. 
]{
\ifdraftmode
\includegraphics[width=0.45\textwidth]{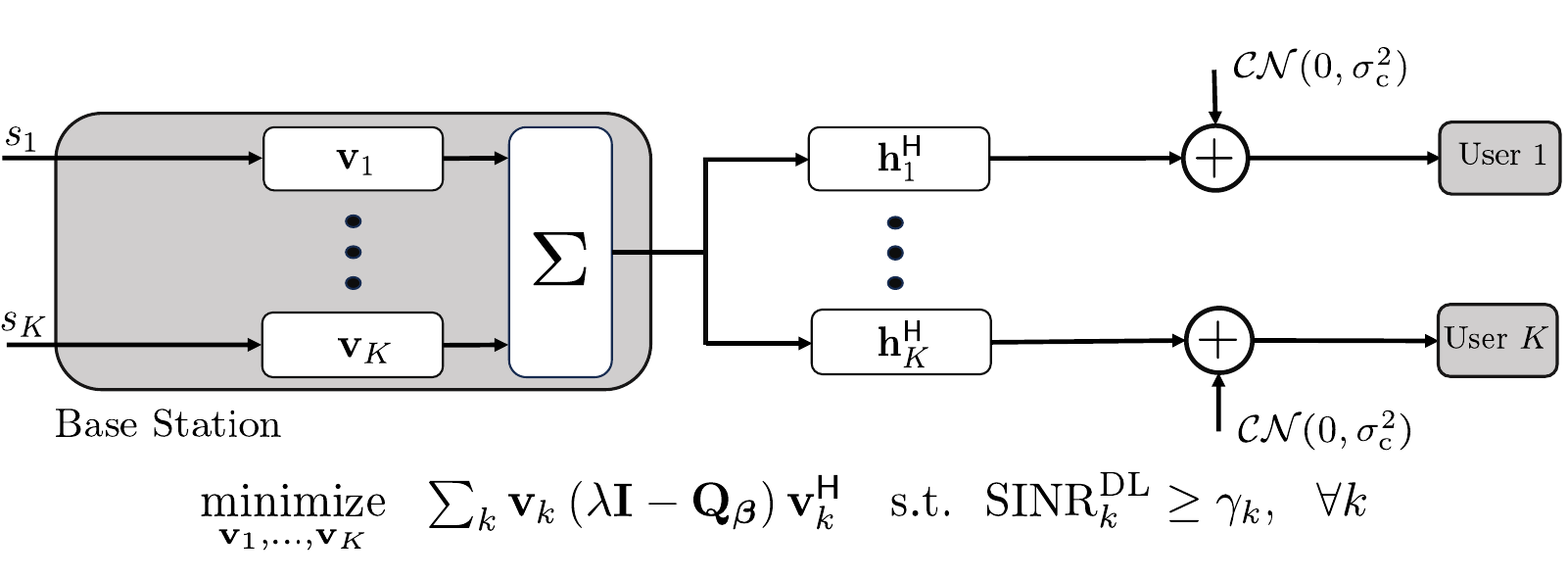}
\else
\includegraphics[width=0.475\textwidth]{figures/DL_c.pdf}
\fi
}
\quad
\subfloat[Uplink beamforming and power minimization problem with noise ``covariance'' 
 $(\lambda \mathbf{I} - \mathbf{Q}_{\boldsymbol{\beta}})$ for satisfying the uplink SINR constraints.
]{
\ifdraftmode
\includegraphics[width=0.45\textwidth]{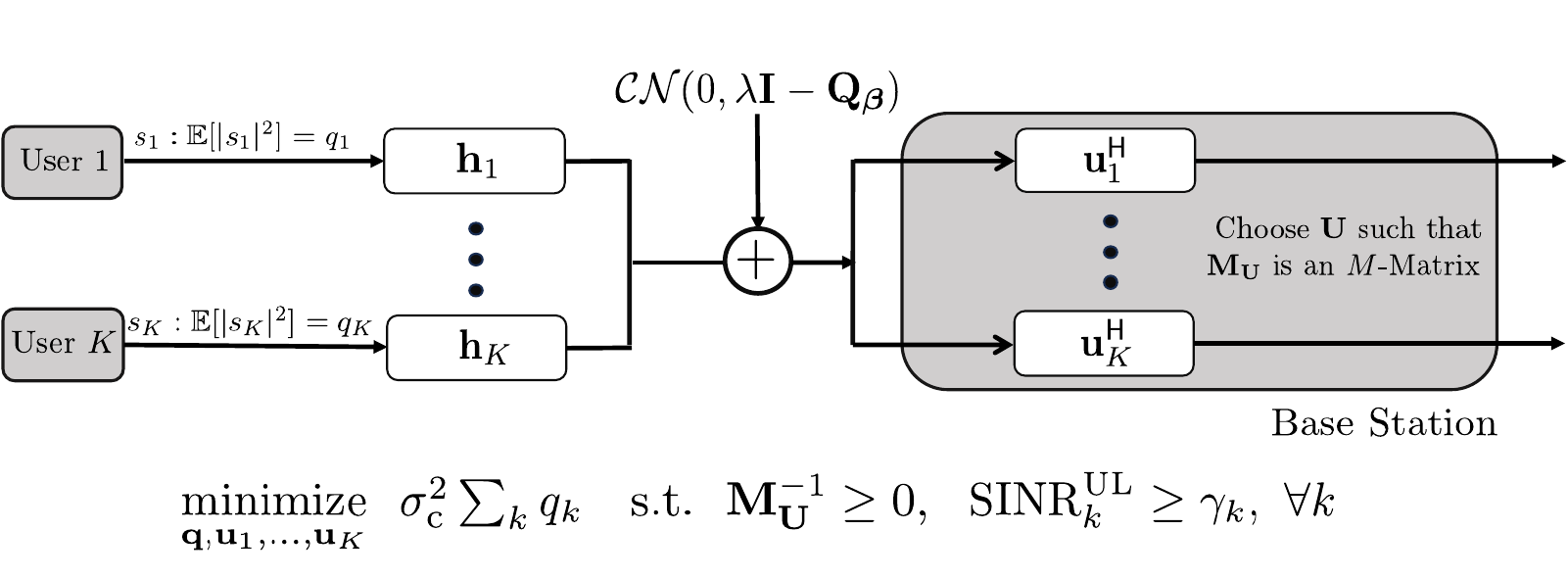}
\else
\includegraphics[width=0.475\textwidth]{figures/UL_c.pdf}
\fi
}
\caption{Uplink-downlink duality holds for downlink weighted power minimization with non-PSD $(\boldsymbol\lambda \mathbf{I} - \mathbf{Q}_\beta)$, if $(\lambda, \boldsymbol{\beta})$ is admissible. 
	The normalized optimal beamformers $\mathbf{v}_k$ for the downlink (left) are the same as the optimal beamformers $\mathbf{u}_k$ for the uplink (right) with an $M$-matrix constraint.
	}
\label{fig:ULDL_duality}
\end{figure*}

Based on the preceding discussion, the statement of Theorem~\ref{thm:ULDL_duality} is that, when $(\lambda, \boldsymbol{\beta})$ is admissible, the optimal beamformers for the uplink problem~\eqref{prob:UL_problem_fixed_lambda} are also optimal as downlink beamforming directions for the downlink problem~\eqref{prob:DL} and vice versa.
Furthermore, the two problems have the same optimal value.
\figurename~\ref{fig:ULDL_duality} illustrates the duality between the downlink and uplink systems that gives rise to~\eqref{prob:DL} and~\eqref{prob:UL_problem_fixed_lambda}. We now proceed to prove the theorem.

\begin{IEEEproof}[Proof of Theorem~\ref{thm:ULDL_duality}]
We first prove that the optimal value of the problem~\eqref{prob:DL} is bounded below if and only if $(\lambda, \boldsymbol{\beta}) \in \mathcal{A}$. Suppose $(\lambda, \boldsymbol{\beta}) \in \mathcal{A}$, then by definition of admissibility, 
for any $\mathbf{V}$ satisfying the SINR constraints, the weighted downlink power is bounded below by zero. Consequently, the optimal value must also be bounded below by zero. Conversely, suppose $(\lambda, \boldsymbol{\beta}) \not \in \mathcal{A}$. Then, there exists some $\tilde{\mathbf{V}}$ satisfying the SINR constraints with $\sum_k \tilde{\mathbf{v}}_k^\textsf{H} \left( \lambda \mathbf{I} - \mathbf{Q}_{\boldsymbol{\beta}} \right)\tilde{\mathbf{v}}_k  < 0$. In this case, it can be easily checked that the scaled beamformers $c \tilde{\mathbf{V}}$ with $c > 1$ continue to satisfy the SINR constraints. Taking $c \rightarrow \infty$ makes the optimal value unbounded below.
   
Now consider the downlink problem~\eqref{prob:DL}, but over 
$\mathbf{p} \triangleq \left[p_1, \ldots, p_K\right]^\mathsf{T} \in \mathbb{R}^{K}$ and $\mathbf{U} \triangleq \left[\mathbf{u}_1, \ldots, \mathbf{u}_K\right] \in \mathbb{C}^{N_\text{T} \times K}$  with
\begin{equation}~\label{eq:upfromv}
    \mathbf{u}_k = \frac{\mathbf{v}_k}{\|\mathbf{v}_k\|}, \quad p_k = \|\mathbf{v}_k\|^2, ~~\forall k
\end{equation}
i.e., $\mathbf{p}$ is a vector of downlink powers and $\mathbf{U}$ is a matrix of normalized beamforming directions. Define the vector ${\boldsymbol \omega}_{\mathbf{U}} \in \mathbb{R}^K$ as 
\begin{equation} \label{def:omega}
\left[{\boldsymbol \omega}_{\mathbf{U}} \right]_k \triangleq \mathbf{u}_k^\textsf{H} (\lambda \mathbf{I} - \mathbf{Q}_{\boldsymbol{\beta}})\mathbf{u}_k.
\end{equation}
Then, we can rewrite
problem~\eqref{prob:DL} as a double minimization 
\begin{subequations} \label{prob:DL_in_U_p}
\begin{align}
    \underset{\mathbf{U}}{\mathrm{minimize}} \quad \underset{\mathbf{p}}{\min}~~~~& 
\boldsymbol{\omega}_{\mathbf{U}}^\mathsf{T} \mathbf{p} \label{eq:DL_weights} \\
    \mathrm{s.t.}~~~~& \mathbf{M}_{\mathbf U} \mathbf{p} \geq \sigma_\text{c}^2 \mathbf{1}, \label{eq:DL_M} \\
& \mathbf{p} \geq 0, \label{eq:DL_p} \\
& \| \mathbf{u}_k \| = 1, ~ \forall k \label{eq:DL_u_normal}
\end{align}
\end{subequations}
where (\ref{eq:DL_M}) comes from an algebraic manipulation of the SINR constraints
(\ref{prob:DL_sinr}). 

The matrix $\mathbf{M}_{\mathbf U}$ has the specific form of having positive
diagonal elements and having off-diagonal elements that are less than or equal to zero. 
We claim that as long as $\mathbf{U}$, along with some $\mathbf{p}$, is a feasible solution to (\ref{prob:DL_in_U_p}),
the matrix $\mathbf{M}_{\mathbf U}$ must belong to a class of so-called \emph{M-matrices} \cite{plemmons1977m}. 
There are many equivalent definitions of the $M$-matrix; several of these are listed in
Appendix \ref{appen:mmat}.
The reason that $\mathbf{M}_{\mathbf U}$ must be 
an $M$-matrix is because the feasibility of (\ref{eq:DL_M})-(\ref{eq:DL_p}) implies the existence of a $\mathbf{p} \ge 0 $ for which $\mathbf{M}_{\mathbf U} \mathbf{p} > 0$. This is one of the equivalent definitions of $M$-matrix.

The converse to the above statement is also true, i.e., if some $\mathbf{U}$ consisting
of normalized beamformers 
is such that $\mathbf{M}_{\mathbf{U}}$ is an $M$-matrix, 
then such $\mathbf{U}$ along with some $\mathbf{p}$ is a feasible solution to 
(\ref{prob:DL_in_U_p}). The reason such feasible $\mathbf{p}$ can be found is that an $M$-matrix is invertible, and the inverse has nonnegative entries. 
So, 
we can obtain a feasible $\mathbf{p}$ by setting (\ref{eq:DL_M}) to equality, i.e., 
\begin{equation}
\label{eq:p_inverse_M}
\mathbf{p} = \sigma_\text{c}^2 \mathbf{M}_{\mathbf{U}}^{-1} \mathbf{1},
\end{equation} 
which, by the aforementioned property of an $M$-matrix, is always nonnegative, so it automatically satisfies (\ref{eq:DL_p}).
Thus, $\mathbf{U}$ being feasible is equivalent to $\mathbf{M}_{\mathbf{U}}$ being an $M$-matrix. 

We further claim that for any feasible $\mathbf{U}$, the above solution for $\mathbf{p}$ is actually an optimal solution to the inner minimization problem of (\ref{prob:DL_in_U_p}).
This would have been a straightforward statement 
if $ (\lambda \mathbf{I} - \mathbf{Q}_{\boldsymbol{\beta}})$ is PSD, in which case
the objective of weighted power minimization (\ref{eq:DL_weights}) would have 
nonnegative weights, i.e., 
$[{\boldsymbol \omega}_{\mathbf{U}}]_k \ge 0$, $\forall k$, due to
\eqref{def:omega}.
But for this downlink ISAC problem, the weights $\boldsymbol{\omega}_{\mathbf U}$ 
are \emph{not necessarily nonnegative}. 
This is because $ (\lambda \mathbf{I} - \mathbf{Q}_{\boldsymbol{\beta}}) $ is not necessarily PSD.  
Furthermore, the admissibility 
condition (\ref{eq:quad_sum}) only ensures that the sum power (or equivalently 
$\boldsymbol{\omega}_{\mathbf{U}}^\mathsf{T} \mathbf{p}$) is nonnegative, but 
allows the individual weights to potentially be negative. This makes sense for 
sensing problems, because the negative weights indicate 
the benefit of \emph{increasing} power toward the target directions.

Nevertheless, we can use the condition $(\lambda, \boldsymbol{\beta}) \in \mathcal{A}$ to
show that for any feasible $\mathbf{U}$, the inner minimization in (\ref{prob:DL_in_U_p}) 
would always have an optimal solution achieving equality in (\ref{eq:DL_M}), and thus 
an optimal $\mathbf{p}$ is given by (\ref{eq:p_inverse_M}). To see this, we make
a change of variable and define
\begin{equation}
\mathbf{\tilde{p}} \triangleq \mathbf{M}_{\mathbf U} \mathbf{p}.
\end{equation}
Then, the inner minimization in (\ref{prob:DL_in_U_p}) can be rewritten as
\begin{subequations} \label{prob:DL_in_U_p_tilde}
\begin{align}
    \underset{\mathbf{\tilde p}}{\mathrm{minimize}}~~~~& 
\boldsymbol{\omega}_{\mathbf{U}}^\mathsf{T} \mathbf{M}_{\mathbf{U}}^{-1} \mathbf{\tilde p} \label{eq:DL_weights_tilde} \\
    \mathrm{subject \; to}  ~~~&  \mathbf{\tilde p} \geq \sigma_\text{c}^2 \mathbf{1}, \label{eq:DL_M_tilde} \\
& \mathbf{M}_{\mathbf{U}}^{-1} \mathbf{\tilde p} \geq 0. \label{eq:DL_p_tilde} 
\end{align}
\end{subequations}
We state the following lemma.
\begin{lemma}
\label{lemma:positive}
Let $(\lambda, \boldsymbol{\beta}) \in \mathcal{A}$. 
Let $\mathbf{U}$ be a matrix of normalized beamformers.
Let $\mathbf{M}_\mathbf{U}$, $\boldsymbol \omega_{\mathbf U}$ be defined as in (\ref{eq:M}) and (\ref{def:omega}) respectively. 
Suppose that $\mathbf{M}_{\mathbf{U}}$ is an $M$-matrix. Then, 
\begin{equation}\label{eq:cond_2}
     \boldsymbol{\phi}^\mathsf{T}_{\mathbf{U}} \triangleq 
\boldsymbol{\omega}_{\mathbf{U}}^\mathsf{T} \mathbf{M}_{\mathbf{U}}^{-1} 
\geq 0.
\end{equation}
\end{lemma}
\begin{IEEEproof}
By definition, $(\lambda, \boldsymbol{\beta}) \in \mathcal{A}$ 
means that the total downlink power is always nonnegative for any feasible 
$\mathbf{U}$ and $\mathbf{p}$ satisfying the downlink SINR constraints.
Since the objective of the problem~\eqref{prob:DL_in_U_p_tilde} is a
reformulation of the objective of the downlink beamforming problem
\eqref{prob:DL_in_U_p} or equivalently \eqref{prob:DL}, and $\mathbf{U}$ 
along with some $\mathbf{p}$ being feasible is equivalent to
$\mathbf{M}_{\mathbf{U}}$ being an $M$-matrix, we can state the
admissibility condition~\eqref{eq:quad_sum} in terms of the transformed
variables $\mathbf{U}$ and $\tilde{\mathbf{p}}$ in the following way:
$(\lambda, \boldsymbol{\beta}) \in \mathcal{A}$ if and only if the
optimal value of (\ref{prob:DL_in_U_p_tilde}) is always nonnegative for all
$\mathbf{U}$ consisting of normalized beamformers such that $\mathbf{M}_{\mathbf{U}}$ 
is an $M$-matrix. 

But this is possible only if the coefficients of $\mathbf{\tilde p}$ in the objective function \eqref{eq:DL_weights_tilde}, i.e.,
$\boldsymbol{\phi}^\mathsf{T}_{\mathbf{U}} \triangleq \boldsymbol{\omega}_{\mathbf{U}}^\mathsf{T} \mathbf{M}_{\mathbf{U}}^{-1}$, are all non-negative.
This is because of the following. First, problem (\ref{prob:DL_in_U_p_tilde}) is clearly 
feasible. This is because we can set $\mathbf{\tilde{p}} = \mathbf{1}$, 
which satisfies (\ref{eq:DL_M_tilde}); further, the fact that $\mathbf{M}_{\mathbf{U}}$ 
is an $M$-matrix ensures (\ref{eq:DL_p_tilde}) is also satisfied. 
Now, if one of $[\boldsymbol{\phi}_\mathbf{U}]_i$ is negative, then we can arbitrarily increase the corresponding $\mathbf{\tilde p}_i$ to drive the objective 
function \eqref{eq:DL_weights_tilde} to negative without
violating the constraints. This is because increasing the value of any $\mathbf{\tilde p}_i$
clearly does not violate (\ref{eq:DL_M_tilde}); further, since $\mathbf{M}_{\mathbf{U}}^{-1}$ is the inverse
of an $M$-matrix, it has nonnegative entries, so increasing $\mathbf{\tilde p}_i$ 
also does not violate the constraint (\ref{eq:DL_p_tilde}). 
Since the optimal value of (\ref{prob:DL_in_U_p_tilde}) cannot be negative
due to the asssumption $(\lambda, \boldsymbol{\beta}) \in \mathcal{A}$,
we must have $ \boldsymbol{\phi}^\mathsf{T}_{\mathbf{U}} 
\geq 0$.
\end{IEEEproof}

Now, let $(\lambda, \boldsymbol{\beta}) \in \mathcal{A}$. For any feasible $\mathbf{U}$ (or equivalently $\mathbf{M}_\mathbf{U}$ is an $M$-matrix), 
we have $\boldsymbol{\phi}_\mathbf{U} \ge 0$ by Lemma~\ref{lemma:positive}.
Then, we argue that there must be one optimal solution of
(\ref{prob:DL_in_U_p_tilde}) that achieves equality in (\ref{eq:DL_M_tilde}). This is 
because if the inequality is strict for some $\mathbf{\tilde p}_i$, one can always decrease 
$\mathbf{\tilde p}_i$ without violating (\ref{eq:DL_p_tilde}) (since $\mathbf{M}_\mathbf{U}^{-1} \ge 0$), while improving the objective (\ref{eq:DL_weights_tilde}) 
or at least keeping it
the same (since $\boldsymbol{\phi}^\mathsf{T}_{\mathbf{U}} \ge 0$). 
This shows that $\mathbf{\tilde p} = \sigma_\text{c}^2 \mathbf{1}$ is an optimal solution, so an optimal
$\mathbf{p}$ is given by (\ref{eq:p_inverse_M}). Note that in the case where $\boldsymbol{\phi}_\mathbf{U}$ has zero entries, multiple optimal solutions for $\mathbf{p}$ exist. They all give the same optimal value. However, one of them is always given by~\eqref{eq:p_inverse_M}. 

Plugging this optimal solution $\mathbf{p}$ into the overall 
downlink beamforming problem (\ref{prob:DL_in_U_p}), we conclude that 
when $(\lambda, \boldsymbol{\beta}) \in \mathcal{A}$, the downlink problem 
(\ref{prob:DL}) is equivalent to
\begin{subequations} \label{prob:DL_in_U}
\begin{align}
\underset{\mathbf{U}}{\mathrm{minimize}}~~~~& 
\sigma_\text{c}^2 \boldsymbol{\omega}_{\mathbf{U}}^\mathsf{T} \mathbf{M}_{\mathbf{U}}^{-1} \mathbf{1} \\
    \mathrm{subject \; to}  ~~~& \mathbf{M}_\mathbf{U}^{-1} \ge 0,  \label{eq:DL_M_matrix} \\
& \| \mathbf{u}_k \| = 1, ~ \forall k \label{eq:DL_u_normal_final}
\end{align}
\end{subequations}
where (\ref{eq:DL_M_matrix}) states that $ \mathbf{M}_\mathbf{U}$ is 
an $M$-matrix. This is needed because otherwise 
a feasible $\mathbf{p}$ would not exist, so such $\mathbf{U}$ cannot be the optimal solution of the downlink. 


Next, we analyze the uplink problem (\ref{prob:UL_problem_fixed_lambda}).
The idea is also to express the uplink problem as a double minimization
\begin{subequations} \label{prob:UL_in_U_p}
\begin{align}
    \underset{\mathbf{U}}{\mathrm{minimize}} \quad \underset{\mathbf{q}}{\min}~~~~& 
\sigma_\text{c}^2 
\mathbf{1}^\mathsf{T} \mathbf{q} \label{eq:UL_weights} \\
    \mathrm{s.t.} ~~~~& \mathbf{M}_{\mathbf U}^\mathsf{T} \mathbf{q} \geq 
	\boldsymbol{\omega}_{\mathbf{U}} \label{eq:UL_M} \\
	& \mathbf{q} \geq 0, \quad \| \mathbf{u}_k \| = 1, ~ \forall k \label{eq:UL_u_normal} \\
	& \mathbf{M}_\mathbf{U}^{-1} \ge 0.  \label{eq:UL_M_matrix} 
\end{align}
\end{subequations}
For the uplink problem, we choose to explicitly include the requirement that 
$\mathbf{M}_\mathbf{U}$ is an $M$-matrix as in constraint~\eqref{eq:UL_M_matrix}
to match the downlink formulation.
Note that this extra constraint would not have been needed if the uplink noise covariance  $(\lambda
\mathbf{I} - \mathbf{Q}_{\boldsymbol{\beta}}) \succ 0$,
because in that case we have $\boldsymbol{\omega}_{\mathbf U} > 0$, due to definition \eqref{def:omega},
and the existence of a feasible power allocation $\mathbf{q}$ satisfying
\eqref{eq:UL_M}-\eqref{eq:UL_u_normal}
implies that $\mathbf{M}_\mathbf{U}$ is an $M$-matrix.

As compared to the downlink,
the definitions of $\mathbf{M}_{\mathbf U}$ and $\boldsymbol{\omega}_{\mathbf{U}}$ here are exactly the 
same, but the SINR constraints are now expressed in terms of 
$\mathbf{M}_{\mathbf U}^\mathsf{T}$. 
Further, the weights in the power minimization and the noises have their roles reversed. 

We now argue that the solution to the inner minimization problem above 
is achieved when the uplink SINRs (i.e.,~\eqref{eq:ul_sinr} or~\eqref{eq:UL_M}) are met with equality, just as in the downlink. 
This is because for every $\mathbf{U}$ and $\mathbf{q}$ satisfying (\ref{eq:UL_M}) and (\ref{eq:UL_M_matrix}) we must have 
$\mathbf{q} \geq 
\mathbf{M}_{\mathbf U}^\mathsf{-T}
	\boldsymbol{\omega}_{\mathbf{U}}$. 
By Lemma \ref{lemma:positive}, since we assume admissibility, we have that $\mathbf{M}_{\mathbf U}^\mathsf{-T}
	\boldsymbol{\omega}_{\mathbf{U}}$  must be nonnegative. Now, since the weights in the objective are all positive (i.e., the all-one vector), we conclude that the optimal  solution to the inner minimization problem must be 
\begin{equation}
\label{eq:q_inverse_M}
\mathbf{q} = \mathbf{M}_{\mathbf{U}}^\mathsf{-T} \boldsymbol{\omega}_\mathbf{U}.
\end{equation}
As an side note, the above solution for $\mathbf{q}$ implies that~\eqref{eq:ult_sinr_opt_eq} holds, because for $\mathbf{U} = \mathbf{U}^*_{\lambda, \boldsymbol{\beta}}$, \eqref{eq:q_inverse_M} becomes $\mathbf{M}^\mathsf{T}_{\mathbf{U}^*_{\lambda, \boldsymbol{\beta}}}\mathbf{q}^* = \boldsymbol{\omega}_{\mathbf{U}^*_{\lambda, \boldsymbol{\beta}}}$, which is exactly~\eqref{eq:ult_sinr_opt_eq}.

Plugging the structure~\eqref{eq:q_inverse_M} into~\eqref{prob:UL_in_U_p}, the uplink problem can now be reduced to 
\begin{subequations} \label{prob:UL_in_U}
\begin{align}
\underset{\mathbf{U}}{\mathrm{minimize}}~~~~& 
\sigma_\text{c}^2 \mathbf{1}^\mathsf{T} \mathbf{M}_{\mathbf{U}}^\mathsf{-T} \boldsymbol{\omega}_{\mathbf{U}} \\
    \mathrm{subject \; to}  ~~~& \mathbf{M}_\mathbf{U}^{-1} \ge 0,  \label{eq:UL_M_matrix_final} \\
& \| \mathbf{u}_k \| = 1, ~ \forall k. \label{eq:UL_u_normal_final}
\end{align}
\end{subequations}
Clearly, (\ref{prob:UL_in_U}) is the same as (\ref{prob:DL_in_U}). 
By noticing that (\ref{eq:optimal_p}) can be obtained from (\ref{eq:p_inverse_M}), and (\ref{eq:opt_V_fixed}) is the same as (\ref{eq:upfromv}),
this proves the stated uplink-downlink duality.
\end{IEEEproof}

The key difference between this new uplink-downlink duality relationship and
the classical duality theory is that the uplink noise covariance  $(\lambda
\mathbf{I} - \mathbf{Q}_{\boldsymbol{\beta}})$ here is not necessarily PSD. 
Thus, there are potentially choices of $\mathbf{u}_k$ for which the uplink noise powers are 
negative. Theorem \ref{thm:ULDL_duality} states that when the
additional $M$-matrix constraint is imposed on the uplink beamformers, the
uplink-downlink duality continues to hold \emph{despite} the potentially negative uplink
noise powers. 

Appendix~\ref{Appen:M-matrix} presents a numerical example that
shows when the $M$-matrix constraint is omitted from the uplink formulation, 
there can be choices of $\mathbf{u}_k$'s that give rise to feasible solutions 
for the uplink, but not for the downlink. In this case, the uplink-downlink
duality breaks down. 

It should be emphasized that the uplink-downlink duality relationship holds only 
for admissible $(\lambda, \boldsymbol{\beta})$. To illustrate the importance of this admissibility
condition, we return to the AoA estimation example.


\subsection{AoA Estimation Example}

Returning to the example of estimating the AoA while communicating with $K$ users. By incorporating the power constraint into the objective in~\eqref{prob:max_min_AoA_final_form}, we obtain the following max-min formulation
\begin{subequations}\label{eq:AoA_DL_prob}
\begin{align}
    \underset{b\in \mathbb{C}, \lambda \geq 0}{\mathrm{maximize}}~\min_{\mathbf{V}} &~\Tr\left( \left( \lambda\mathbf{I} - \mathbf{Q}_b \right) \mathbf{V} \mathbf{V}^\textsf{H} \right) 
   - \frac{|b|^2}{\sigma_\alpha^2} -\lambda P \\
   \mathrm{s. t.} &~~  \frac{\left|\mathbf{h}_k^\textsf{H} \mathbf{v}_k \right|^2}{\sum_{i \neq k} \left|\mathbf{h}_k^\textsf{H} \mathbf{v}_i \right|^2 + \sigma_\text{c}^2} \geq \gamma_k, \quad \forall k
\end{align}
\end{subequations}
For fixed $(\lambda, b)$, the inner problem can be viewed as a downlink communications problem similar to the classical problem~\eqref{prob:dl_comm}, provided that the pair $(\lambda, b)$ is admissible. 

To check admissibility, we note that
the inner problem~\eqref{eq:AoA_DL_prob} is bounded if and only if its dual problem (which is a semidefinite program) is feasible. 
This provides a numerical way of characterizing the admissible set by checking whether the dual problem is feasible for different values of $(\lambda, b)$. 
     
\begin{figure}
    \centering
    \includegraphics[width=0.273\textwidth]{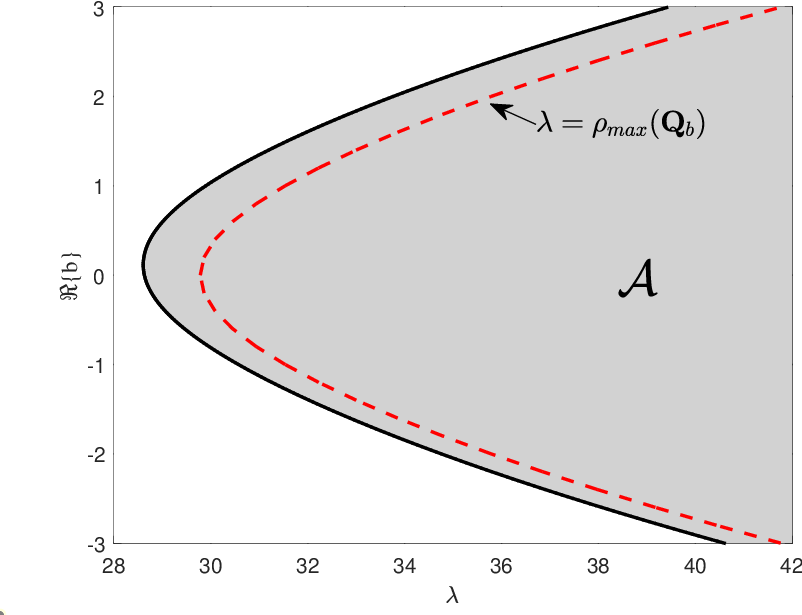}
    \caption{A cut of the admissible set when $\Im{\left\{b\right\}} = 1$ for the AoA estimation example with $N_\text{T} = N_\text{R} = 16$, $P = 10$, 
$K = 4$ users with SINR constraints set to be $8, 10, 10$ and $12$dB.} 
    \label{fig:admissible_set}
\end{figure}
 
In~\figurename~\ref{fig:admissible_set}, we show a cut in the admissible set obtained at fixed $\Im\{b\} = 1$ across different values of $\lambda$ and $\Re\{b\}$ for a random realization of the channels, where $\alpha \sim \mathcal{CN}(1,1)$ for the sensing channel, and $h_{ij} \sim \mathcal{CN}(0,1)$ for the communication channel. In the figure, the admissible set $\mathcal A$ is given by the shaded region. 
This is to be contrasted with the set of $(\lambda, b)$ values for which $(\lambda \mathbf{I} - \mathbf{Q}_b)$ is PSD, given by the set of $(\lambda, b)$ values to the right of the dashed curve obtained by setting $\lambda$ to be the largest eigenvalue of $\mathbf{Q}_b$. 

Theorem~\ref{thm:ULDL_duality} states that for any $(\lambda, b)$ outside the admissible set $\mathcal{A}$, the inner minimization in~\eqref{eq:AoA_DL_prob} is unbounded. Moreover, for any $(\lambda, b)$ in the admissible region, the inner problem admits an equivalent formulation as an uplink problem with $(\lambda \mathbf{I} - \mathbf{Q}_b)$ acting as the uplink noise covariance matrix. 
As seen from the figure, the admissible set $\mathcal{A}$ is strictly larger than the set of $(\lambda, b)$ for which $(\lambda \mathbf{I} - \mathbf{Q}_b)$ is PSD. 
This underscores the need to develop the new uplink-downlink duality theory for the case where  $(\lambda \mathbf{I} - \mathbf{Q}_b)$ is admissible but not PSD\footnote{If we assume that $b$ is fixed and consider only $\lambda$ for the outer optimization, the admissibility set reduces to a ray, as in \cite{jie2014} for the energy harvesting scenario. Thus, the formulation in this paper encompasses~\cite{jie2014} as a special case.  However, the analysis in \cite{jie2014} appears to have overlooked the $M$-matrix constraint for the uplink beamformers and the ensuing complication associated with negative noise power in the duality formulation.}.

    

\section{Numerical Algorithm and Simulations}\label{sec:numerical}


The uplink-downlink duality is useful in devising low-complexity algorithms
for solving 
the ISAC problem~\eqref{prob:generalCase}. 
The strategy
involves solving the downlink problem~\eqref{prob:DL} for each fixed 
choice of the pair $(\lambda, \boldsymbol{\beta})$, using its equivalent 
uplink formulation~\eqref{prob:UL_problem_fixed_lambda} in an iterative fashion in an inner loop, followed by 
updating $\left(\lambda, \boldsymbol{\beta}\right)$ in an outer loop. 

However, in contrast to the classical uplink-downlink duality for the communication-only scenario, the duality relation for the ISAC problem only holds for the $(\lambda, \boldsymbol{\beta})$ pairs that are admissible. \editrev{Thus, the uplink-downlink duality algorithm for ISAC must include a check on admissibility.} Furthermore, the uplink problem~\eqref{prob:UL_problem_fixed_lambda} contains an additional $M$-matrix constraint to ensure the uplink beamformers lead to a feasible solution for the downlink problem. Thus, the iterations must ensure that the $M$-matrix constraint is satisfied.



Despite such complications, it turns out that a modified version of the classical uplink-downlink algorithm can be used for the ISAC problem in the inner loop. Furthermore, when combined with a proper outer loop for updating $(\lambda, \boldsymbol{\beta})$, the optimal solution of the ISAC problem~\eqref{prob:generalCase} can be obtained under a set of sufficient conditions. \editrev{The algorithmic details, including beamformer design, power control, convergence analysis, proper initialization, admissibility check, and the condition under which the optimal solution to the ISAC problem can be recovered, are rather involved and will be deferred to future work.} Here, we use the AoA estimation example to illustrate the optimized beam patterns obtained from the algorithm. 

Consider the recurring example of estimating the AoA for a single target with $N_\text{T} = N_\text{R} = 20$ and $P = 10$.  For the sensing target, we consider a uniform prior for $\theta$ over the interval $\left[-\Delta, \Delta\right]$, where $\Delta$ is a parameter that determines the range of $\theta$ in the prior. The path loss coefficient $\alpha$ is assumed to follow $\mathcal{CN}(1, 1)$ a priori. 
To avoid the logarithm of the prior to go to minus infinity at the edge of the uniform distribution, a common practice is to approximate the prior by a smooth function or by a Gaussian prior with the same mean and variance for computing the matrix $\mathbf{C}$ in \eqref{eq:c_elements}; see~\cite{Huleihel2013optimal}. Here, we adopt the latter approach for simplicity. 

\begin{figure}
    \centering    \includegraphics[width=0.4\textwidth]{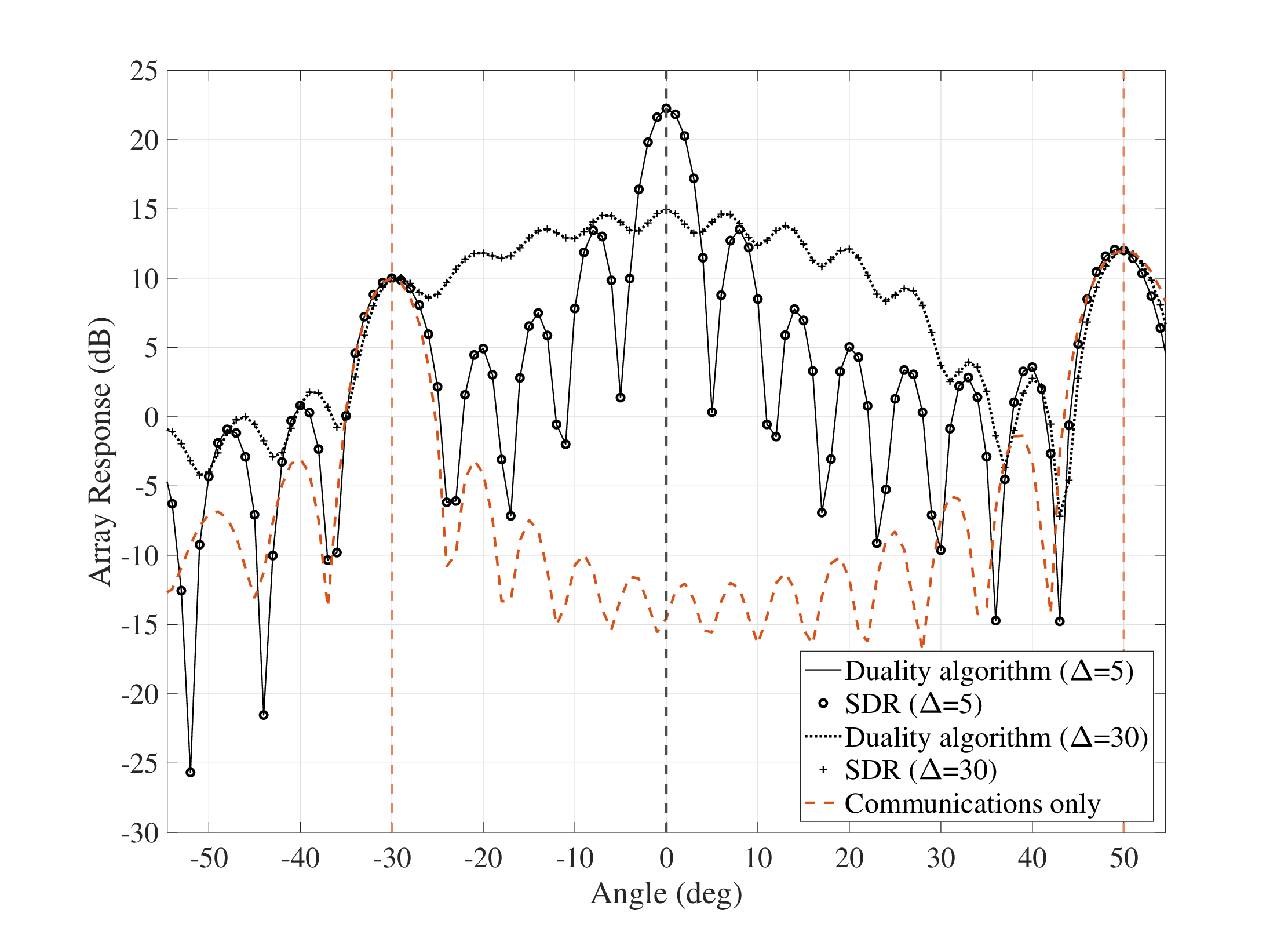}
    \caption{Beam patterns of uplink-downlink duality solution vs. SDR solution. 
The communication users are at -30$^\circ$ and 50$^\circ$. The sensing target is uniformly distributed at $[-\Delta,\Delta]$ with $\Delta = 5^\circ$ or $30^\circ$.
}
     \label{fig:comparison}
\end{figure}

For communications, we set $K = 2$ users to ensure that Assumption~\ref{main_assume} holds. \editrev{The SINR thresholds are set to be $10$ and $12$ dB. We assume that the communication channels also have a single LoS path with angles at $-30^\circ$ and $50^\circ$ and gain equal to one.} 


\figurename~\ref{fig:comparison} compares the beam pattern of the SDR solution with the beam pattern obtained using the uplink-downlink duality algorithm. The comparison is shown for both narrow and wide priors with $\Delta = 5^\circ$ and $30^\circ$, respectively. In both cases, it is seen that the beam patterns obtained from both algorithms match exactly. Numerically, the optimal values of the objective function also match. 
Both algorithms achieve a global optimum. \editrev{Note that the users have unequal beam amplitudes due to the fact that their SINR requirements are different.}


The figure shows that the optimal beamforming strategy forms beams in the directions of both the communication users and the sensing target, but interestingly, a wider beam pattern is utilized when $\Delta$ is larger in order to account for the more uncertainty in estimating $\theta$ in that case.  We also plot the beam pattern associated with the optimal solution of the classical communication problem~\eqref{prob:dl_comm} 
without sensing. We see that the resulting beam pattern focuses towards the communication users, but is clearly oblivious of any sensing target, as expected. 

\section{Conclusion}\label{sec:conc}
This paper presents an uplink-downlink duality theory for beamforming design in
ISAC systems.  We focus on systems for which the use of communication beams alone
already attains optimal performance for both communications and sensing, and
establish a duality relationship between the downlink beamforming problem and 
a \emph{virtual} uplink problem in which the noise covariance matrix is not
necessarily PSD. The non-PSD uplink noise covariance arises because in
contrast to the conventional downlink beamforming problem for communication
alone, the ISAC system aims to \emph{maximize} the beamformed power in the sensing
directions of interest rather than simply to minimize the transmit power. 
This paper shows that the sensing directions of interest can be characterized
precisely via an analysis of the CRB corresponding to the underlying parameter
estimation problem.  Moreover, the paper establishes an \emph{admissibility}
condition under which the uplink-downlink duality relationship for this new
setting holds, and furthermore reveals an additional \emph{M-matrix} condition
that must be included on the virtual uplink beamformers in the duality relationship.
These theoretical insights open doors for the development of numerical
algorithms for beamformer design that are much more efficient than the
conventional SDR approach for downlink beamforming for the ISAC problem.




\ifdraftmode
\bibliographystyle{IEEEtran}
\bibliography{IEEEabrv,ref}
\fi

%

%








\ifdraftmode
\newpage
{
\centering
SUPPLEMENTARY MATERIALS \\
}
\fi

\appendices

\section{Proof of Lemma~\ref{lem:SDR_tight}}
\label{Appen:A}

Assumption~\ref{main_assume} states that the optimal value of problem~\eqref{prob:generalCase} is equal to the optimal value of the following ISAC optimization that would arise if one instead employs the extended model~\eqref{eq:BFMDL_extended}:
\begin{subequations}\label{prob:general_NT_BF}
    \begin{align}
    \underset{\mathbf{\tilde V} \in \mathbb{C}^{N_\text{T} \times (N_\text{T} + K)}}{\mathrm{minimize}} ~ & \Tr \left(\mathbf{W} \mathbf{J}^{-1}_\mathbf{\tilde V}  \right)  \\
            \mathrm{subject \ to}  ~~ & \frac{\left|\mathbf{h}_k^\textsf{H} \mathbf{v}_k \right|^2}{\sum_{i \neq k} \left|\mathbf{h}_k^\textsf{H} \mathbf{v}_i \right|^2 + \mathbf{h}_k^\textsf{H} \mathbf{V}_\text{s}^\mathsf{H}  \mathbf{V}_\text{s}\mathbf{h}_k + \sigma_\text{c}^2} \geq \gamma_k, \ \ \forall k  \\
            & \| \mathbf{\tilde V} \|_\text{F}^2 \leq P
    \end{align}
\end{subequations}
where $\mathbf{\tilde V} = \left[ \begin{matrix} \mathbf{V} & \mathbf{V}_\text{s}  \end{matrix} \right] \in \mathbb{C}^{N_\text{T} \times (N_\text{T} + K)}$ is an extended beamforming matrix with a matrix of communication beamformers $\mathbf{V} = \left[\mathbf{v}_1, \ldots, \mathbf{v}_K\right] \in \mathbb{C}^{N_\text{T} \times K}$ and a matrix of sensing beamformers $\mathbf{V}_\text{s} \in \mathbb{C}^{N_\text{T} \times N_\text{T}}$. So, to prove the desired result that problem~\eqref{prob:generalCase} has a tight SDR, we only need to prove that i) problem~\eqref{prob:general_NT_BF} has a tight SDR; and ii) the SDRs of problems~\eqref{prob:generalCase} and~\eqref{prob:general_NT_BF} are equivalent.

The proof of the first claim follows immediately from a standard trick used in several ISAC studies that has been used to show that problems with a similar structure have tight SDR; see, e.g.,~\cite{ Liu2020joint, LiuFCRB2022}. We provide the details here to keep this paper self-contained. 

Consider the SDR of problem~\eqref{prob:general_NT_BF}, which can be obtained by defining $\mathbf{R} = \mathbf{ \tilde V} \mathbf{ \tilde V}^\mathsf{H}$, and $\mathbf{R}_k = \mathbf{v}_k \mathbf{v}_k^\mathsf{H}$ for all $k$, then dropping the rank-one constraints on $\mathbf{R}_1, \ldots, \mathbf{R}_K$:
\begin{subequations}\label{prob:general_NT_BF_SDR}
    \begin{align}
\underset{\mathbf{ R}, \mathbf{R}_1,\ldots, \mathbf{R}_K}{\mathrm{minimize}} ~~&  \Tr \left(\mathbf{W} \mathbf{J}_{\mathbf{ R}}^{-1} \right) 
     \\
    \mathrm{subject \ to}   ~& \left(1 + \frac{1}{\gamma_k}\right)  \mathbf{h}_k^\textsf{H} \mathbf{R}_k \mathbf{h}_k -   \mathbf{h}_k^\textsf{H} \mathbf{R} \mathbf{h}_k \geq \sigma_\text{c}^2, \ \ \forall k  \label{eq:sinr_const_appen} \\
& \Tr \left( \mathbf{R} \right) \leq P, \\
& \mathbf{   R} \succcurlyeq \sum_k \mathbf{R}_k, 
\quad \mathbf{R}_k \succcurlyeq 0 ~~\forall k. \label{eq:dummy_const_appen}
    \end{align}
\end{subequations}
To show that the above SDR is a tight relaxation of problem~\eqref{prob:general_NT_BF}, we show the existence of a rank-one optimal solution. Let $\mathbf{\hat R}, \mathbf{\hat  R}_1, \ldots, \mathbf{\hat R}_K$ denote an arbitrary high rank solution of~\eqref{prob:general_NT_BF_SDR}. We claim that $\mathbf{R}^*, \mathbf{R}_1^*, \ldots, \mathbf{R}_K^*$ as defined below is also a solution:
\begin{equation}\label{eq:rank_one_solution}
    \mathbf{R}^* = \mathbf{\hat{R}}, \quad \mathbf{R}_k^* = \frac{\mathbf{\hat{R}}_k \mathbf{h}_k^\mathsf{H} \mathbf{h}_k \mathbf{\hat R}_k }{\mathbf{h}_k^\mathsf{H}  \mathbf{\hat R}_k\mathbf{h}_k}, \quad \forall k.
\end{equation}
Note that by construction, it is clear that $\mathbf{R}^*_1, \ldots \mathbf{R}^*_K$ are rank-one PSD matrices. 

Moreover, this set of $\mathbf{R}^*, \mathbf{R}_1^*, \ldots, \mathbf{R}_K^*$ satisfies the power constraint and attains the same objective value as $\mathbf{\hat{R}}, \mathbf{\hat{R}}_1, \ldots, \mathbf{\hat{R}}_K$, because the overall covariance is kept the same, i.e., $ \mathbf{R}^* = \mathbf{\hat R}$. 

Next, we note that
\begin{align}\label{eq:rank_one_conditions2}
\mathbf{h}_k^\mathsf{H}\mathbf{\hat{R}}_k \mathbf{h}_k  = \mathbf{h}_k^\mathsf{H}\mathbf{R}^*_k \mathbf{h}_k, \quad \forall k
\end{align}
which is immediate from~\eqref{eq:rank_one_solution}, so that
\begin{align}
&\left(1 + \frac{1}{\gamma_k} \right) \mathbf{h}_k^\mathsf{H}\mathbf{R}^*_k \mathbf{h}_k -\mathbf{h}_k^\mathsf{H}\mathbf{R}^* \mathbf{h}_k  \nonumber \\
&= \left(1 + \frac{1}{\gamma_k} \right) \mathbf{h}_k^\mathsf{H}\mathbf{\hat R}_k \mathbf{h}_k - \mathbf{h}_k^\mathsf{H}\mathbf{\hat R} \mathbf{h}_k \geq \sigma_\text{c}^2.
\end{align}
In other words, the SINR constraints~\eqref{eq:sinr_const_appen} are also satisfied.

Finally, for an arbitrary vector $\mathbf{w} \in \mathbb{C}^{N_\text{T}}$, the following is true due to Cauchy-Schwarz inequality
\begin{equation}
    \mathbf{w}^\mathsf{H} \left( \mathbf{\hat{R}}_k - \mathbf{R}^*_k\right) \mathbf{w} = \mathbf{w}^\mathsf{H} \mathbf{\hat{R}}_k \mathbf{w} - \frac{\left| \mathbf{h}_k^\mathsf{H} \mathbf{\hat{R}}_k \mathbf{w} \right|^2}{\mathbf{h}_k^\mathsf{H} \mathbf{\hat{R}}_k \mathbf{h}_k } \geq 0.
\end{equation}
This shows that $\mathbf{w}^\mathsf{H} \mathbf{\hat{R}}_k\mathbf{w} \geq \mathbf{w}^\mathsf{H}   \mathbf{R}^*_k\mathbf{w}$, $\forall \mathbf{w}$, or equivalently $\mathbf{\hat R}_k \succcurlyeq \mathbf{R}^*_k$, which implies that
\begin{equation}
    \mathbf{R}^* = \mathbf{\hat R}  \succcurlyeq \sum_k \mathbf{\hat{R}}_k \succcurlyeq \sum_k \mathbf{R}^*_k.
\end{equation}
In other words, the constraint \eqref{eq:dummy_const_appen} is also satisfied.
This shows that the SDR~\eqref{prob:general_NT_BF_SDR} has a rank-one solution and that the relaxation is tight.

Next, we aim to prove that the SDR of problem~\eqref{prob:generalCase} (as given in~\eqref{prob:general_SDP_main_text}), and the SDR problem~\eqref{prob:general_NT_BF_SDR} are equivalent. To do so, it suffices to show that any feasible solution of one problem can be mapped into a feasible solution for the other problem. Let $\mathbf{\tilde R}_1, \ldots, \mathbf{\tilde R}_K$ be an arbitrary feasible solution of the SDR~\eqref{prob:general_SDP_main_text}. Then, it is easily verified that the solution
$\mathbf{R}, \mathbf{R}_1, \ldots, \mathbf{R}_K$ defined below satisfies all constraints in~\eqref{prob:general_NT_BF_SDR}
\begin{equation}
    \mathbf{R} = \sum \mathbf{\tilde R}_k, \quad \mathbf{R}_k = \mathbf{\tilde R}_k, \quad \forall k.
\end{equation}

Conversely, suppose that $\mathbf{R}, \mathbf{R}_1, \ldots, \mathbf{R}_K$ is a feasible solution of problem~\eqref{prob:general_NT_BF_SDR}. Then, we claim that
\begin{equation}\label{eq:inverse_construction}
    \mathbf{\tilde R}_k = \mathbf{R}_k + \frac{1}{K} \left( \mathbf{R} - \sum_i \mathbf{R}_i \right), \quad \forall k
\end{equation}
is a feasible solution of~\eqref{prob:general_SDP_main_text}. To see this, we note that
\begin{equation}\label{eq:proof_inequality}
    \mathbf{\tilde R}_k \succcurlyeq \mathbf{R}_k \succcurlyeq 0, \quad  \forall k
\end{equation} since the second term in~\eqref{eq:inverse_construction} is PSD as given by~\eqref{eq:dummy_const_appen}. 
Next, we note that
\begin{subequations}\label{eq:construct_sum}
\begin{align}
    \sum_k \mathbf{\tilde R}_k  = \sum_k \left(\mathbf{R}_k + \frac{1}{K} \left( \mathbf{R} - \sum_i \mathbf{R}_i \right) \right) \\
    = \left(\sum_k \mathbf{R}_k\right) + \mathbf{R} - \sum_i \mathbf{R}_i  = \mathbf{R} 
\end{align}
\end{subequations}
i.e., the matrices $\mathbf{\tilde R}_k$ in~\eqref{eq:inverse_construction} sum up to $\mathbf{R}$. Therefore, the power constraint in~\eqref{prob:general_SDP_main_text} is satified for $\mathbf{\tilde{R}}_1, \ldots, \mathbf{\tilde{R}}_K$, i.e., 
\begin{equation}
    \Tr\left (\sum_k \mathbf{\tilde R}_k \right) = \Tr( \mathbf{R} ) \leq P.
\end{equation}
Finally, the SINR constraints in~\eqref{prob:general_SDP_main_text} (i.e., constraints~\eqref{eq:sdp_const_1})  are also satisfied due to the following
\begin{subequations}
\begin{align}
    \left(1 + \frac{1}{\gamma_k}\right)  \mathbf{h}_k^\textsf{H} \mathbf{\tilde R}_k \mathbf{h}_k &\geq \left(1 + \frac{1}{\gamma_k}\right)  \mathbf{h}_k^\textsf{H} \mathbf{R}_k \mathbf{h}_k \\
    &\geq  \mathbf{h}_k^\textsf{H} \mathbf{R} \mathbf{h}_k + \sigma_\text{c}^2 \\
    &= \sum_k \mathbf{h}_k^\textsf{H} \mathbf{\tilde R}_k \mathbf{h}_k + \sigma_\text{c}^2
\end{align}
\end{subequations}
where the first line follows from~\eqref{eq:proof_inequality}, the second line follows from the SINR constraint~\eqref{eq:sinr_const_appen}, and the third line follows from~\eqref{eq:construct_sum}. This shows that SDRs~\eqref{prob:general_SDP_main_text} and~\eqref{prob:general_NT_BF_SDR} are equivalent. 

Combining the above with Assumption~\ref{main_assume} and the fact that the relaxation~\eqref{prob:general_NT_BF_SDR} is tight, we conclude that all four problems (i.e., problem~\eqref{prob:generalCase}, problem~\eqref{prob:general_NT_BF}, and their SDRs) are equivalent. 

\section{Equivalence of Problem~\eqref{prob:max_min_special_case} and Problem~\eqref{prob:max_min_AoA_final_form}}\label{appen:simple_minimax_formulation_AoA}

First, recall that we assume that $\alpha$ and $\theta$ are independent with Gaussian priors so that 
\begin{equation}
\mathbf{C} = \frac{2T}{\sigma_\text{s}^2} \diag\left(\frac{1}{\sigma^2_\alpha}, \frac{1}{\sigma^2_\alpha}, \frac{1}{\sigma^2_\theta}\right) \succ 0.
\end{equation}
It follows that $\mathbf{J}_\mathbf{V} \succ 0$ in~\eqref{eq:BFIM_AoA}, because $\mathbf{J}_\mathbf{V} = \mathbf{C} + \mathbf{T}_\mathbf{V}$ and $\mathbf{T}_\mathbf{V} \succcurlyeq 0$. 

Under Assumption~\ref{main_assume}, we can interchange the $\max$ and the $\min$ in problem~\eqref{prob:max_min_special_case} to obtain the equivalent problem
 \begin{align} 
\underset{\mathbf{V} \in \mathcal{V}}{\mathrm{minimize}}~\max_{b_1, b_2, b_3} ~& 2 b_3 -\boldsymbol{\beta}_3^\mathsf{T} \mathbf{C} \boldsymbol{\beta}_3 - \Tr{\left(\mathbf{Q}_{\boldsymbol{\beta}_3} \mathbf{V}\mathbf{V}^H \right)}.
\end{align}
We now claim that the optimal $b_3$ in the inner problem above always satisfy
$b_3 > 0$. This is because we can use \eqref{eq:beta_relation}, \eqref{eq:D-Qrelation} and  $b_3 = \mathbf{e}_3^\mathsf{T} \boldsymbol{\beta}_3$ 
to rewrite the inner problem as 
\begin{align}\label{prob:max_beta_AoA}
\underset{\boldsymbol{\beta}_3 \in \mathbb{R}^3}{\mathrm{maximize}} ~~2 \mathbf{e}_3^\mathsf{T} \boldsymbol{\beta}_3 - \boldsymbol{\beta}_3^\mathsf{T} \mathbf{J}_\mathbf{V} \boldsymbol{\beta}_3,
\end{align}
which has a solution $\boldsymbol{\beta}_3^* = \mathbf{J}^{-1}_\mathbf{V} \mathbf{e}_3$. Noting that $ \mathbf{J}_\mathbf{V} \succ 0$, this implies 
\begin{equation}\label{eq:positive_b3}
    b_3^* = \mathbf{e}_3^\mathsf{T} \boldsymbol{\beta}_3^* = \mathbf{e}_3^\mathsf{T} \mathbf{J}^{-1}_\mathbf{V} \mathbf{e}_3 > 0.
\end{equation}
Thus, we can restrict to $b_3 > 0$ without loss of optimality.

Now, consider the following change of variables:
\begin{equation}\label{eq:map}
    a \triangleq b_3 \in \mathbb{R}_{++}, \quad b \triangleq \tfrac{1}{b_3} (b_1 + \imath b_2) \in \mathbb{C},
\end{equation}
which is related to the perspective function (see~\cite{boyd2004convex}).

Since $b_3 \neq 0$, the mapping is well defined. 
Furthermore, given $(a,b)$ with $a>0$, we can uniquely recover $b_1 = \Re\{ab\}$, $b_2 = \Im\{ab\}$, and $b_3 = a$. In fact, the mapping from $(b_1,b_2,b_3) \in \mathbb{R}^2 \times \mathbb{R}_{++}$ to $(a,b) \in \mathbb{R}_{++} \times \mathbb{C}$ is bijective. 
This means that instead of optimizing over $(b_1,b_2,b_3)$, we can 
reformulate the problem as an optimization over $(a, b)$. 

Now, it can be verified that
\begin{subequations}
\begin{align}
\boldsymbol{\beta}_3^\mathsf{T}\mathbf{C}\boldsymbol{\beta}_3 &= \frac{2T}{\sigma_\text{s}^2 \sigma^2_\alpha}\left(b_1^2 + b_2^2\right) + \frac{2T}{\sigma_\text{s}^2 \sigma^2_\theta} b^2_3 \\
&= \frac{2T}{\sigma_\text{s}^2} a^2  \left( \frac{|b|^2}{\sigma_\alpha^2} + \frac{1}{\sigma^2_\theta} \right),
\end{align}
\end{subequations}
and $\tilde{\mathbf{G}}^{(\alpha,\theta)}_{\boldsymbol{\beta}} =  (b_1 + \imath b_2) \mathbf{A}(\theta)  + b_3 \alpha \dot{\mathbf{A}}(\theta) =  a (b \mathbf{A}(\theta)  + \alpha \dot{\mathbf{A}}(\theta))$, so that
\begin{subequations}
\begin{align}
\mathbf{Q}_{\boldsymbol{\beta}_3} &= \frac{2T}{\sigma_\text{s}^2} \mathbb{E}_{\alpha, \theta}\left[ \left(\tilde{\mathbf{G}}^{(\alpha,\theta)}_{\boldsymbol{\beta}_3}\right)^\textsf{H} \tilde{\mathbf{G}}^{(\alpha,\theta)}_{\boldsymbol{\beta}_3} \right] \\
&= \frac{2T}{\sigma_\text{s}^2}  a^2 \mathbb{E}_{\alpha, \theta}\left[ \left(b \mathbf{A}(\theta)  + \alpha \dot{\mathbf{A}}(\theta)\right)^\textsf{H} \left(b \mathbf{A}(\theta)  + \alpha \dot{\mathbf{A}}(\theta)\right) \right] \\
&= \frac{2T}{\sigma_\text{s}^2}  a^2 \mathbf{Q}_b.
\end{align}
\end{subequations}
where $\mathbf{Q}_b$ is defined in~\eqref{prob:max_min_AoA_final_form_Qb}. 
Then, problem~\eqref{prob:max_min_special_case} can be stated as
\begin{align} 
\label{prob:mini_max_AoA2}
\underset{\mathbf{V} \in \mathcal{V}}{\mathrm{minimize}} \max_{a>0, b} 2 a - \frac{2T}{\sigma_\text{s}^2}  a^2 
\left(  \frac{|b|^2}{\sigma_\alpha^2} + \frac{1}{\sigma^2_\theta} + \Tr\left(\mathbf{Q}_b \mathbf{V} \mathbf{V}^\mathsf{H} \right) \right) 
\end{align}
For fixed $\mathbf{V}$ and $b$, the maximization over $a$ occurs at
\begin{equation}\label{eq:op_a}
    a^* = \frac{1}{\frac{2T}{\sigma_\text{s}^2} 
\left(  \frac{|b|^2}{\sigma_\alpha^2} + \frac{1}{\sigma^2_\theta} + \Tr\left(\mathbf{Q}_b \mathbf{V} \mathbf{V}^\mathsf{H} \right) \right)} > 0.
\end{equation}
Substituting back into~\eqref{prob:mini_max_AoA2} and taking the reciprocal yield 
\begin{align} \label{prob:mini_max_final_form}
\underset{\mathbf{V} \in \mathcal{V}}{\mathrm{maximize}}~\min_{b \in \mathbb{C}} &~~ \frac{2T}{\sigma_\text{s}^2}  
\left(  \frac{|b|^2}{\sigma_\alpha^2} + \frac{1}{\sigma^2_\theta} + \Tr\left(\mathbf{Q}_b \mathbf{V} \mathbf{V}^\mathsf{H} \right) \right),
\end{align}
As final step, we eliminate the constants $\tfrac{2T}{\sigma_\text{s}^2}$ and $\frac{1}{\sigma^2_\theta}$, and again interchange max and min to arrive at the desired form \eqref{prob:max_min_AoA_final_form}. 

It turns out that the alternative formulation~\eqref{prob:max_min_AoA_final_form} can also be derived by analyzing the dual of the following equivalent formulation of problem~\eqref{prob:simpleCase}:
\begin{subequations}\label{prob:simpleCase_Schur}
\begin{align}
\underset{\mathbf{V} \in \mathcal{V}}{\mathrm{maximize}} ~~&  d' \\
\mathrm{subject \; to }
~~& \left[\begin{matrix} h_\theta(\mathbf{V}) - d' & h_{\alpha\theta}(\mathbf{V}) \\  \bar{h}_{\alpha\theta}(\mathbf{V}) & h_\alpha(\mathbf{V}) + \tfrac{1}{\sigma_\alpha^2} \end{matrix} \right] \succcurlyeq 0. \label{eq:sd_const_aoa}
\end{align}
\end{subequations}
This can be accomplished by dualizing with respect to constraint~\eqref{eq:sd_const_aoa} and by following the same steps as in the proof of Theorem~\ref{thm:max_min}. 

Let $\mathbf{\tilde B}$ denote the $2\times 2$ dual matrix associated with~\eqref{eq:sd_const_aoa}:
\begin{equation}
    \mathbf{\tilde B} = \left[\begin{matrix}
        b' & \bar{b} \\
        b & b''
    \end{matrix} \right] \succcurlyeq 0
\end{equation}
where $b' \in \mathbb{R}$, $b \in \mathbb{C}$, and $b'' \in \mathbb{R}$. 
The Lagrangian dual of the problem \eqref{prob:simpleCase_Schur} 
is
\begin{align*}
\underset{\mathbf{\tilde B} \succcurlyeq 0}{\mathrm{minimize}}\max_{\mathbf{V} \in \mathcal{V}, d'}   d' (1 - b') + b' h_\theta(\mathbf{V}) + \bar{b} \bar{h}_{\alpha \theta}(\mathbf{V}) \\
~~~~+ b h_{\alpha \theta}(\mathbf{V})
     + b'' \left(h_\alpha(\mathbf{V}) + \tfrac{1}{\sigma^2_\alpha}\right)
    \end{align*}
which is equivalent to problem~\eqref{prob:simpleCase} due to strong duality (i.e., under Assumption~\ref{main_assume}). In the above problem, we must have $b' = 1$ since otherwise the inner optimization is unbounded. 
Substituting $b' = 1$, we get   
\begin{align}\label{eq:max_min_simpleCase}
\underset{b'' \geq |b|^2}{\mathrm{minimize}}~\max_{ \mathbf{V} \in \mathcal{V} }    h_\theta(\mathbf{V}) + \bar{b} \bar{h}_{\alpha \theta}(\mathbf{V}) +  b h_{\alpha \theta}(\mathbf{V}) \nonumber \\
     + b'' \left(h_\alpha(\mathbf{V}) + \tfrac{1}{\sigma^2_\alpha}\right)
    \end{align}
 where the condition $b'' \ge |b|^2 $ follows from
    $\tilde{\mathbf{B}} \succcurlyeq 0$ and $b' = 1$, so the Schur complement must satisfy $b'' - |b|^2 \ge 0$. 

By interchanging min and max, it is easy to see that the optimal $b''$ is attained when $b'' = |b|^2$, since $(h_\alpha(\mathbf{V}) + \tfrac{1}{\sigma^2_\alpha})$ is positive for every $\mathbf{V}$.
Substituting this into \eqref{eq:max_min_simpleCase}, we arrive at
\begin{align}\label{eq:max_min_simpleCase_v2}
\underset{b \in \mathbb{C}}{\mathrm{minimize}}~\max_{\mathbf{V} \in \mathcal{V}}   ~~ h_\theta(\mathbf{V}) + \bar{b} \bar{h}_{\alpha \theta}(\mathbf{V}) +  b h_{\alpha \theta}(\mathbf{V}) \nonumber \\
     + |b|^2 h_\alpha(\mathbf{V}) + \tfrac{|b|^2}{\sigma^2_\alpha}
\end{align}
Finally, it can be verified that:
\begin{equation}
    h_\theta(\mathbf{V}) + \bar{b} \bar{h}_{\alpha \theta}(\mathbf{V}) +  b h_{\alpha \theta}(\mathbf{V}) \nonumber \\
     + |b|^2 h_\alpha(\mathbf{V}) = \Tr\left(\mathbf{Q}_b \mathbf{V} \mathbf{V}^\mathsf{H} \right),
\end{equation}
so that problem~\eqref{eq:max_min_simpleCase_v2} is precisely problem~\eqref{prob:max_min_AoA_final_form}.

\section{Brief Review of $M$-Matrices}\label{appen:mmat}
A matrix $\mathbf{B}$ is termed an $M$-matrix if two conditions hold:
\begin{itemize}
    \item The diagonal elements are nonnegative but the off-diagonal elements are nonpositive, i.e., 
    \begin{equation}
         [\mathbf{B}]_{ii} \geq 0, ~\forall i. \quad \text{and}  \quad  
         [\mathbf{B}]_{ij} \leq 0, ~\forall i \neq j,
    \end{equation}
    \item The inverse exists and has nonnegative elements, i.e., $[\mathbf{B}^{-1}]_{ij} \geq 0, \forall i, j$.
\end{itemize}
$M$-matrices play an important role in stability analysis in Control Theory. For this reason, they have been extensively studied. There exist several equivalent characterizations of the second condition, as long as the first condition above holds. Of interest to this paper are the following equivalent conditions:
\begin{enumerate}
    \item $\mathbf{B}^{-1}$ exists and is nonnegative.
    \item There exists a vector $\mathbf{p} \geq 0$ satisfying $\mathbf{B} \mathbf{p} > 0$.
    \item $\mathbf{B} = \mathbf{D} - \mathbf{F}$ with $\mathbf{D}^{-1} \:{\geq}~ 0$, $\mathbf{F} \geq 0$ and $\rho_\text{max}(\mathbf{D}^{-1} \mathbf{F})<1$.
\end{enumerate}
A proof of equivalence of these conditions, as well as an extensive list of equivalent characterizations can be found in~\cite{plemmons1977m}.

\section{Necessity of $M$-Matrix Constraint in the Dual Uplink Formulation: An Example}
\label{Appen:M-matrix}
We use a numerical example to demonstrate that when 
$\left(\lambda \mathbf{I} - \mathbf{Q}_{\boldsymbol{\beta}}\right) $ is not PSD, 
the constraint that $\mathbf{M}_{\mathbf U}$ is an $M$-matrix must be included in 
the uplink formulation~\eqref{prob:UL_problem_fixed_lambda} for the uplink-downlink
duality relationship to hold. 
This is so even for admissible $(\lambda, \boldsymbol{\beta})$. 

Let $N_\text{T} = N_\text{R} = 2$ and $K = 2$ with the following channel realizations
\begin{equation}
    \mathbf{h}_1 = \left[ \begin{matrix} 1 \\ 0 \end{matrix} \right], \quad \mathbf{h}_2 = \left[ \begin{matrix} 0 \\ 1 \end{matrix} \right],
\end{equation}
and the following SINR thresholds and noise power:
\begin{equation}
    \gamma_1 = 4, \quad \gamma_2 = 2, \quad \sigma_\text{c}^2 = 1.
\end{equation}
For sensing, we consider the AoA estimation problem with independent $\alpha$ and $\theta$. The priors are $\alpha \sim \mathcal{CN}(0, 1)$ and $\theta \sim \mathcal{N}(0, 1)$.
Set $b = 1$ as in \eqref{prob:max_min_AoA_final_form_Qb}. 
We can calculate 
\begin{subequations}
\begin{align}\label{eq:Qb1}
	\mathbf{Q}_{b} &= \mathbb{E}_{\alpha, \theta}\left[ ( b \mathbf{A}(\theta) + \alpha \dot{\mathbf{A}}(\theta) )^\textsf{H} ( b \mathbf{A}(\theta) + \alpha \dot{\mathbf{A}}(\theta) ) \right] \\
    &\approx   \left[ \begin{matrix} 1.2566 & -0.0458 \\ -0.0458 & 1.2566 \end{matrix} \right].
\end{align}
\end{subequations}
This matrix has equal diagonal elements and has eigenvectors
\begin{equation}
    \mathbf{w}_1 = \frac{1}{\sqrt{2}} \left[  \begin{matrix}
        1 \\ - 1
    \end{matrix} \right], \quad \mathbf{w}_2 = \frac{1}{\sqrt{2}} \left[  \begin{matrix}
        1 \\ 1
    \end{matrix} \right],
\end{equation}
and eigenvalues
\begin{equation}
    \rho_1 = 1.3024, \quad \rho_2 = 1.2108.
\end{equation}
The set of admissible $(\lambda, b)$'s are those for which the downlink problem is bounded below:
\begin{subequations} \label{eq:downlink_two_users}
\begin{align} 
\underset{\mathbf{v}_1, \mathbf{v}_2}{\mathrm{minimize}}  ~~& \mathbf{v}_1^\mathsf{H} \left(\lambda \mathbf{I} - \mathbf{Q}_b\right) \mathbf{v}_1 + \mathbf{v}_2^\mathsf{H} \left(\lambda \mathbf{I} - \mathbf{Q}_b\right) \mathbf{v}_2  \\
    \mathrm{subject \ to} ~& \frac{ \left|\mathbf{h}_1^\textsf{H} \mathbf{v}_1 \right|^2}{ \left| \mathbf{h}_1^\textsf{H}\mathbf{v}_2\right|^2 +  1}  \geq 4,  \quad \frac{ \left|\mathbf{h}_2^\textsf{H} \mathbf{v}_2 \right|^2}{  \left| \mathbf{h}_2^\textsf{H}\mathbf{v}_2\right|^2 +  1}  \geq  2. \label{eq:downlink_SINRs_two_users}
\end{align}
\end{subequations}  
This problem has strong duality. One way to check whether \eqref{eq:downlink_two_users} is 
bounded below is to check whether 
its dual problem, as given below, is feasible:
\begin{subequations} \label{eq:downlink_dual_two_users}
\begin{align} 
\underset{z_1 \geq 0, z_2 \geq 0}{\mathrm{maximize}}  ~~& z_1 + z_2  \\
    \mathrm{subject \ to} ~~& \lambda \mathbf{I} - \mathbf{Q}_b + z_2 \mathbf{h}_2 \mathbf{h}_2^\mathsf{H} \succcurlyeq \tfrac{z_1}{4} \mathbf{h}_1 \mathbf{h}_1^\mathsf{H}, \label{eq:SDP_feas_const1} \\
    ~~& \lambda \mathbf{I} - \mathbf{Q}_b + z_1 \mathbf{h}_1 \mathbf{h}_1^\mathsf{H} \succcurlyeq \tfrac{z_2}{2} \mathbf{h}_2\mathbf{h}_2^\mathsf{H}. \label{eq:SDP_feas_const3}
\end{align}
\end{subequations} 
Numerically, one can determine that the minimum value of $\lambda$ for which $(\lambda, b)$ is admissible 
is $\lambda_\text{min} \approx 1.297$.

Now, choose 
\begin{equation}
    \lambda = 1.3 > \lambda_\text{min},
\end{equation}
so that the pair $(\lambda, b) = (1.3, 1)$ is admissible. Observe that,
\begin{equation}
    \lambda \mathbf{I} - \mathbf{Q}_{b} = \left[ \begin{matrix}  0.0434 &   0.0458 \\
   0.0458 &   0.0434 \end{matrix} \right]
\end{equation} 
is not PSD since $\lambda < \rho_1$. 

For this $(\lambda \mathbf{I} - \mathbf{Q}_{b})$, we claim that the $M$-matrix constraint cannot be dropped from the uplink formulation, otherwise the duality relationship in Theorem~\ref{thm:ULDL_duality} would not hold. To do this, we first solve the downlink problem~\eqref{eq:downlink_two_users} numerically by SDR. The optimal value of this problem is approximately 
\begin{equation}
    f_\text{DL}^* \approx 0.1435,
\end{equation}
and the corresponding optimal solution is $\mathbf{v}_1^* \approx \left[ \begin{matrix} 2.5048 & -1.1651 \end{matrix} \right]^\mathsf{T}$ and $\mathbf{v}_2^* \approx \left[  \begin{matrix} -0.7540 & 2.1714 \end{matrix} \right]^\mathsf{T}$. After normalizing, we obtain
\begin{equation}
    \mathbf{u}_1^* \approx \left[ \begin{matrix} 0.9067 \\ -0.4218 \end{matrix} \right], \quad \mathbf{u}_2^* \approx \left[  \begin{matrix} -0.3280 \\ 0.9447 \end{matrix} \right] .
\end{equation}
As a first step, let us show that the above $(\mathbf{u}^*_1, \mathbf{u}^*_2)$ attains the same objective value for the uplink problem. In other words, the statement of duality in Theorem~\ref{thm:ULDL_duality} holds for $(\mathbf{u}^*_1, \mathbf{u}^*_2)$. The uplink problem with the $M$-matrix constraint is 
\begin{subequations}
\begin{align} 
\underset{q_1, q_2, \mathbf{u}_1, \mathbf{u}_2}{\mathrm{minimize}}  ~~& q_1 + q_2  \\
    \mathrm{subject \ to} ~& \frac{q_1 \left|\mathbf{h}_1^\textsf{H} \mathbf{u}_1 \right|^2}{ 4}  \geq q_2 \left| \mathbf{h}_2^\textsf{H}\mathbf{u}_1\right|^2 +  \mathbf{u}_1^\textsf{H} 
    \left( \lambda \mathbf{I} - \mathbf{Q}_{b} \right)
    \mathbf{u}_1  \label{eq:UL_const_1}  \\
    & \frac{q_2 \left|\mathbf{h}_2^\textsf{H} \mathbf{u}_2 \right|^2}{ 2}  \geq   q_1 \left| \mathbf{h}_1^\textsf{H}\mathbf{u}_2\right|^2 +  \mathbf{u}_2^\textsf{H} 
    \left( \lambda \mathbf{I} - \mathbf{Q}_{b} \right)
    \mathbf{u}_2 \\
	& q_1 \ge 0, \quad q_2 \geq 0, \label{eq:UL_const_3} \\
    &\mathbf{M}_{\mathbf{U}}^{-1} \geq 0. \label{eq:UL_const_4}
\end{align}
\end{subequations}   
By setting $\mathbf{u}_1 = \mathbf{u}_1^*$ and $\mathbf{u}_2 = \mathbf{u}_2^*$ and computing the noise and the interference terms, we obtain:
\begin{subequations}\label{eq:two_users_M}
\begin{align} 
\underset{q_1, q_2}{\mathrm{minimize}}  ~~& q_1 + q_2  \\
    \mathrm{subject \ to} ~&   \left[ \begin{matrix}
        0.2055 & - 0.1779 \\ -0.1076 & 0.4462
    \end{matrix} \right] \left[ \begin{matrix}
        q_1 \\ q_2
    \end{matrix} \right] \gtrsim \left[ \begin{matrix}
        0.0084 \\ 0.0150 
    \end{matrix} \right] \label{eq:dummy_const_1} \\
	& q_1 \ge 0, \quad q_2 \geq 0. 
\end{align}
\end{subequations} 
Note that the matrix on the right-hand side of~\eqref{eq:dummy_const_1} is $\mathbf{M}_{\mathbf{U}}^\mathsf{T}$, which has the following inverse:
\begin{equation}
    \mathbf{M}_\mathbf{U}^{-T} \approx \left[ \begin{matrix}
        6.1487 &   2.4512 \\
    1.4827   & 2.8322
    \end{matrix} \right] \ge 0.
\end{equation}
In other words, the $M$-matrix constraint~\eqref{eq:UL_const_4} is satified for $\mathbf{u}_1^*, \mathbf{u}_2^*$. In light of this fact, the optimal uplink powers of~\eqref{eq:two_users_M} are obtained when~\eqref{eq:dummy_const_1} is met with an equality. Such uplink powers are readily verified to be $q_1^* \approx 0.0885$ and $q_2^*\approx 0.0551$. The uplink objective value of~\eqref{eq:two_users_M} is:
\begin{equation}
	0.1435 \approx q_1^* + q_2^* = f^*_\text{DL}.
\end{equation}
Thus, the same objective is attained for both uplink and downlink problems, as Theorem~\ref{thm:ULDL_duality} states.

Next, consider the uplink problem as stated below without the $M$-matrix constraint:
\begin{subequations} \label{eq:uplink_two_users}
\begin{align} 
\underset{q_1, q_2, \mathbf{u}_1, \mathbf{u}_2}{\mathrm{minimize}}  ~~& q_1 + q_2  \\
    \mathrm{subject \ to} ~& \eqref{eq:UL_const_1}-\eqref{eq:UL_const_3}
\end{align}
\end{subequations}   
We proceed to show that the optimal value of the above problem is strictly less than $f_\text{DL}^*$.
Set the receive beamformers in the uplink problem to be the eigenvector corresponding to $\rho_1$, i.e., $\mathbf{u}_1 = \mathbf{u}_2 = \mathbf{w}_1$. We next show that this choice is an optimal solution for the uplink problem~\eqref{eq:uplink_two_users}. 

For the above choice of beamformers, the ``virtual" uplink noise powers are \emph{negative}
\begin{equation}\label{eq:u_example}
    \mathbf{u}_1^\textsf{H} 
    \left( \lambda \mathbf{I} - \mathbf{Q}_{b} \right)
    \mathbf{u}_1 =  \mathbf{u}_2^\textsf{H} 
    \left( \lambda \mathbf{I} - \mathbf{Q}_{b} \right)
    \mathbf{u}_2 \approx -0.0024,
\end{equation}
and all the direct and interfering channels are equal, i.e.,
\begin{equation} 
|\mathbf{h}_i^\mathsf{H} \mathbf{u}_j|^2 = \tfrac{1}{2}, \quad \forall i, j = 1, 2.
\end{equation}
With this choice of $({\mathbf u}_1, {\mathbf u}_2)$, the uplink constraints \eqref{eq:UL_const_1}-\eqref{eq:UL_const_3} simplify to:
    \begin{equation}
        \frac{\tfrac{1}{2}q_1}{4 } \gtrsim \tfrac{1}{2}q_2 - 0.0024,
       ~~~ \frac{\tfrac{1}{2}q_2}{2} \gtrsim \tfrac{1}{2}q_1 - 0.0024, ~~~ q_1 \geq 0, ~~~ q_2 \geq 0.
    \end{equation}
It can be readily verified that $q_1' = q_2' = 0$ satisfies the above set of inequalities.
Furthermore, they attain the minimum possible objective value in the uplink, which is strictly less than $f_\text{DL}^*$, i.e.,
\begin{equation}
 q_1' + q_2' = 0 < f_\text{DL}^*.
\end{equation}
In other words, the downlink problem~\eqref{eq:downlink_two_users} and the uplink problem~\eqref{eq:uplink_two_users} have different optimal values. Therefore, uplink-downlink duality does not hold for this uplink problem without the $M$-matrix constraint.

To further confirm the necessity of the $M$-matrix constraint, we demonstrate that the choice $\mathbf{u}_1 = \mathbf{u}_2 = \mathbf{w}_1$ does not give rise to a feasible solution for the downlink problem~\eqref{eq:downlink_two_users}, despite being a feasible choice for the uplink problem~\eqref{eq:uplink_two_users}. 

Set $\mathbf{v}_1 = \sqrt{p_1} \mathbf{w}_1$ and $\mathbf{v}_2 = \sqrt{p_2} \mathbf{w}_1$ as the downlink beamformers in the SINR constraint~\eqref{eq:downlink_SINRs_two_users}. This yields 
\begin{equation}
    \frac{\tfrac{1}{2}p_1 }{\tfrac{1}{2} p_2 + 1} \geq 4, \quad \frac{\tfrac{1}{2} p_2 }{\tfrac{1}{2} p_1 + 1} \geq 2,
\end{equation}
which have no solution for $p_1 \geq 0$ and $p_2 \geq 0$. Indeed, this is because the above constraints can be written as:
\begin{equation}
\left[ \begin{matrix}
    \tfrac{1}{8} & -\tfrac{1}{2} \\ -\tfrac{1}{2} & \tfrac{1}{4} 
\end{matrix} \right] \left[ \begin{matrix}
    p_1 \\ p_2
\end{matrix} \right] \geq \left[ \begin{matrix}
    1 \\ 1
\end{matrix} \right],
\end{equation}
and the matrix on the right-hand side has an inverse with all-negative entries, so it is not an $M$-matrix. Multiplying both sides by the matrix inverse and flipping the inequality, we have
\begin{equation}
    \left[ \begin{matrix}
    p_1 \\ p_2
\end{matrix} \right] \leq \left[ \begin{matrix}
    -\tfrac{24}{7} \\ -\tfrac{20}{7}
\end{matrix} \right].
\end{equation}
Since the power cannot be negative, this choice of beamformers is not feasible for the downlink problem. 

As can be seen in this example, it is the negative noise power in the uplink channel that allows the set of feasible beamformers in the uplink to be larger than the corresponding set of feasible beamformers in the downlink. 
The purpose of the $M$-matrix constraint is to restrict the feasible set of uplink beamformers 
to be exactly those beamformers that are also feasible in the downlink. This is a crucial constraint that cannot be omitted.

\ifdraftmode
\else
\bibliographystyle{IEEEtran}
\bibliography{IEEEabrv,ref}
\fi

\ifCLASSOPTIONcaptionsoff
  \newpage
\fi
\end{document}